\PassOptionsToPackage{numbers,sort&compress}{natbib}  
\documentclass[preprint,12pt]{elsarticle}
\usepackage[margin=2.5cm]{geometry}
\usepackage{enumitem}
\newlist{inlinelist}{enumerate*}{1}
\setlist*[inlinelist,1]{%
  label=(\roman*),
}
\usepackage{dirtytalk}
\usepackage{soul}
\usepackage{hhline}
\usepackage{subfig}
\usepackage{multirow}
\usepackage{xcolor}
\usepackage{graphicx}
\usepackage{upgreek}
\usepackage{amssymb}
\usepackage{amsfonts,amsthm,bm,amsmath} 
\usepackage{appendix}
\usepackage{times}
\usepackage{caption}
\usepackage{subfig}
\newcommand{\psubref}[1]{\protect\subref{#1}}
\usepackage{multirow}
\usepackage{xcolor}
\usepackage[linesnumbered,ruled,vlined]{algorithm2e}
\usepackage{tablefootnote}

\SetCommentSty{mycommfont}

\SetKwInput{KwInput}{Input}                
\SetKwInput{KwOutput}{Output}              

\usepackage[colorlinks]{hyperref} 
\hypersetup{ 
    colorlinks=true,       
    linkcolor=red,          
    citecolor=blue,        
    filecolor=magenta,      
    urlcolor=cyan           
}

\newcommand{\fref}[1]{Fig.~\ref{#1}}

\newcommand{\eref}[1]{Eq.~(\ref{#1})}

\newcommand{\sref}[1]{Section~\ref{#1}}
\newcommand{\tref}[1]{Table~\ref{#1}}
\setcitestyle{square}

\journal{arXiv}
\begin{document}

\begin{frontmatter}

\title{Deep energy method in topology optimization applications}
\author[]{Junyan He$^1$}
\author[]{Shashank Kushwaha$^1$}
\author[]{Charul Chadha$^1$}
\author[]{Seid Koric$^{1,2}$}
\author[]{Diab Abueidda$^2$}
\author[]{Iwona Jasiuk$^1$\corref{mycorrespondingauthor}}
\address{$^1$ Department of Mechanical Science and Engineering, University of Illinois at Urbana-Champaign, Champaign, IL, USA \\
$^2$ National Center for Supercomputing Applications, University of Illinois at Urbana-Champaign, Champaign, IL, USA}
\cortext[mycorrespondingauthor]{Corresponding author}
\ead{ijasiuk@illinois.edu}
\begin{abstract}

This paper explores the possibilities of applying physics-informed neural networks (PINNs) in topology optimization (TO) by introducing a fully self-supervised TO framework that is based on PINNs. This framework solves the forward elasticity problem by the deep energy method (DEM). Instead of training a separate neural network to update the density distribution, we leverage the fact that the compliance minimization problem is self-adjoint to express the element sensitivity directly in terms of the displacement field from the DEM model, and thus no additional neural network is needed for the inverse problem. The method of moving asymptotes is used as the optimizer for updating density distribution. The implementation of Neumann, Dirichlet, and periodic boundary conditions are described in the context of the DEM model. Three numerical examples are presented to demonstrate framework capabilities: (1) Compliance minimization in 2D under different geometries and loading, (2) Compliance minimization in 3D, and (3) Maximization of homogenized shear modulus to design 2D meta material unit cells. The results show that the optimized designs from the DEM-based framework are very comparable to those generated by the finite element method, and shed light on a new way of integrating PINN-based simulation methods into classical computational mechanics problems.

\section*{\bf{Graphical abstract}}
{\centering
\includegraphics[width=0.82\textwidth]{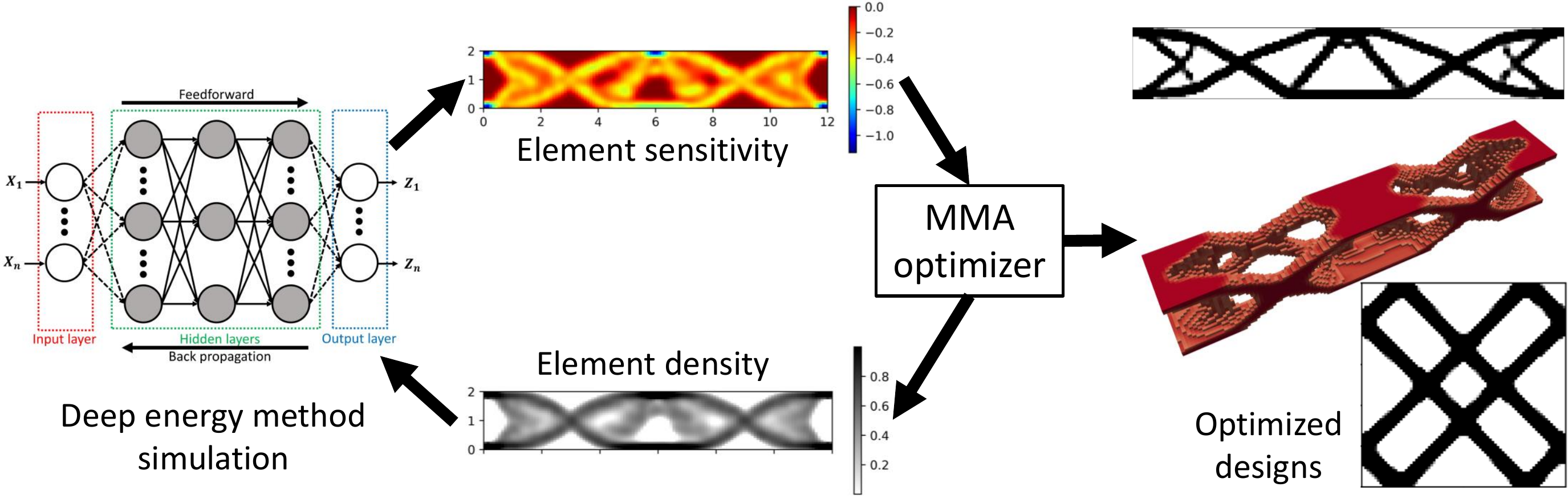}
\par
}
\end{abstract}

\begin{keyword}
Deep energy method \sep Physics-informed neural networks \sep Topology optimization \sep SIMP
\end{keyword}

\end{frontmatter}

\section{Introduction}
\label{sec:intro}
Neural networks (NNs) have seen wide applications in computational mechanics. They have been used to learn the complex deformation process of binary composites \cite{chen2019machine,yang2020prediction,luo2021rapid}, thin-walled shell structures \cite{he2022exploring,chen2019application}, plasticity \cite{stoffel2019neural,gorji2020potential,abueidda2021deep} and to learn constitutive laws \cite{chen2021recurrent,yang2020learning,flaschel2021unsupervised} from a large collection of finite element simulations training set. These models seek to approximate the underlying relations embedded in finite element solutions, but do not seek to solve the underlying partial differential equations (PDEs) representing the governing physics principles. In contrast, physics-informed neural networks (PINNs) \cite{shukla2022scalable,haghighat2021physics,cai2022physics,henkes2022physics,haghighat2021physics} present an alternative to the finite element method (FEM) to numerically solve the governing PDEs directly.

PINNs construct a NN approximation to the PDE solution at discrete points, and the loss function is defined based on how well the underlying physics is satisfied at these training points. One such example of PINN is the deep collocation method \cite{raissi2018deep,abueidda2021meshless,guo2021deep,haghighat2021physics}, which is based directly on the strong form of the governing PDE, and seeks to minimize a loss function defined as the $L_2$ norm of the residual in the PDE at discrete points. Since it is based on the strong form, naturally it involves the computation of second-order (e.g., in elasticity, heat transfer and the Navier-Stokes equations) or higher-order (e.g., plate bending) spatial gradients of the outputs through automatic differentiation of the NN, which can be quite expensive. However, they are relatively straightforward to implement and is applicable to all PDEs, as it mainly involves numerically evaluating the differential operator in the PDE. Therefore, strong-form based PINNs have been applied to solve a wide variety of problems such as linear elasticity, hyperelasticity, plasticity, composite mechanics \cite{yan2022framework,niaki2021physics}, micromechanics \cite{henkes2022physics}, and fluid mechanics \cite{cai2022physics,wessels2020neural}.

An alternative method to avoid calculating high-order gradient is to write the governing equation into a variational form, by finding a functional whose variation yields the governing PDE; this is termed the deep Ritz method \cite{yu2018deep,liao2019deep}. For some physical systems, the governing physics dictate that the solution of the PDE yields the minimum of the energy functional, such as the principle of minimum potential energy in structural mechanics \cite{segerlind1984applied,reddy2014introduction}. In these cases, it is natural to define the energy functional as the loss function, as the training of the NN is a process where the loss value is minimized. This energy-based method is termed the deep energy method (DEM) \cite{samaniego2020energy,nguyen2020deep,abueidda2021deep}. Although DEM avoids working direcly with the strong form of the PDE, it is only limited to physical systems where the solution is given by a minimum potential, which is not applicable in cases like fluid mechanics. Additionally, researchers have proposed and implemented a mixed formulation, where both the strong form and the energy method are used to capture the mechanical response with high solution gradients and stress concentrations \cite{fuhg2022mixed, abueidda2022enhanced, rezaei2022mixed}. 

NNs are also widely used in topology optimization (TO) applications. NNs have been trained on optimized designs generated by finite-element TO simulations to rapidly predict optimized structures for new loading and boundary conditions \cite{kollmann2020deep,abueidda2020topology,sosnovik2019neural,banga20183d}, and are used as a way to parameterize the TO design space as an alternative to the element-based parameterization in FEM-based TO \cite{zhang2021tonr,hoyer2019neural,chandrasekhar2021tounn}. Recently, the work by Zehnder et al. \cite{zehnder2021ntopo} outlined a new way of applying PINNs to TO, where the forward elasticity problem is solved via a DEM-based technique, and another neural network is trained for the inverse design problem to generate density-based gradient and update the design density. This is a fully self-supervised approach, where no inputs from finite-element simulations are needed. Although the training time is longer compared to FEM solution time, the NN-based TO framework generated designs that have similar, if not better performance than FEM-based TO.

Inspired by the work of Zehnder et al. \cite{zehnder2021ntopo}, we introduce a simpler, yet fully self-supervised approach for TO based on the DEM method to shed light on how PINN-based simulation techniques can be combined with traditional solution techniques in computational mechanics to provide novel solutions to classical problems. This paper is organized as follows: \sref{sec:methods} presents an overview of deep neural networks, deep energy method, and topology optimization. \sref{sec:results} presents and discusses the results from three numerical examples. \sref{sec:conc} summarizes the outcomes and limitation of this current study and highlights possible future works.

\section{Methods}
\label{sec:methods}
\subsection{Deep neural networks}
\label{DNN}
Deep neural networks (DNNs) consist of multiple layers of interconnected neurons, and the layers themselves are also connected. The configuration and function of DNNs bear resemblance to that of a human brain. \fref{DNN} shows a typical structure of a fully connected neural network model. 
\begin{figure}[h!] 
    \centering
         \includegraphics[width=0.6\textwidth]{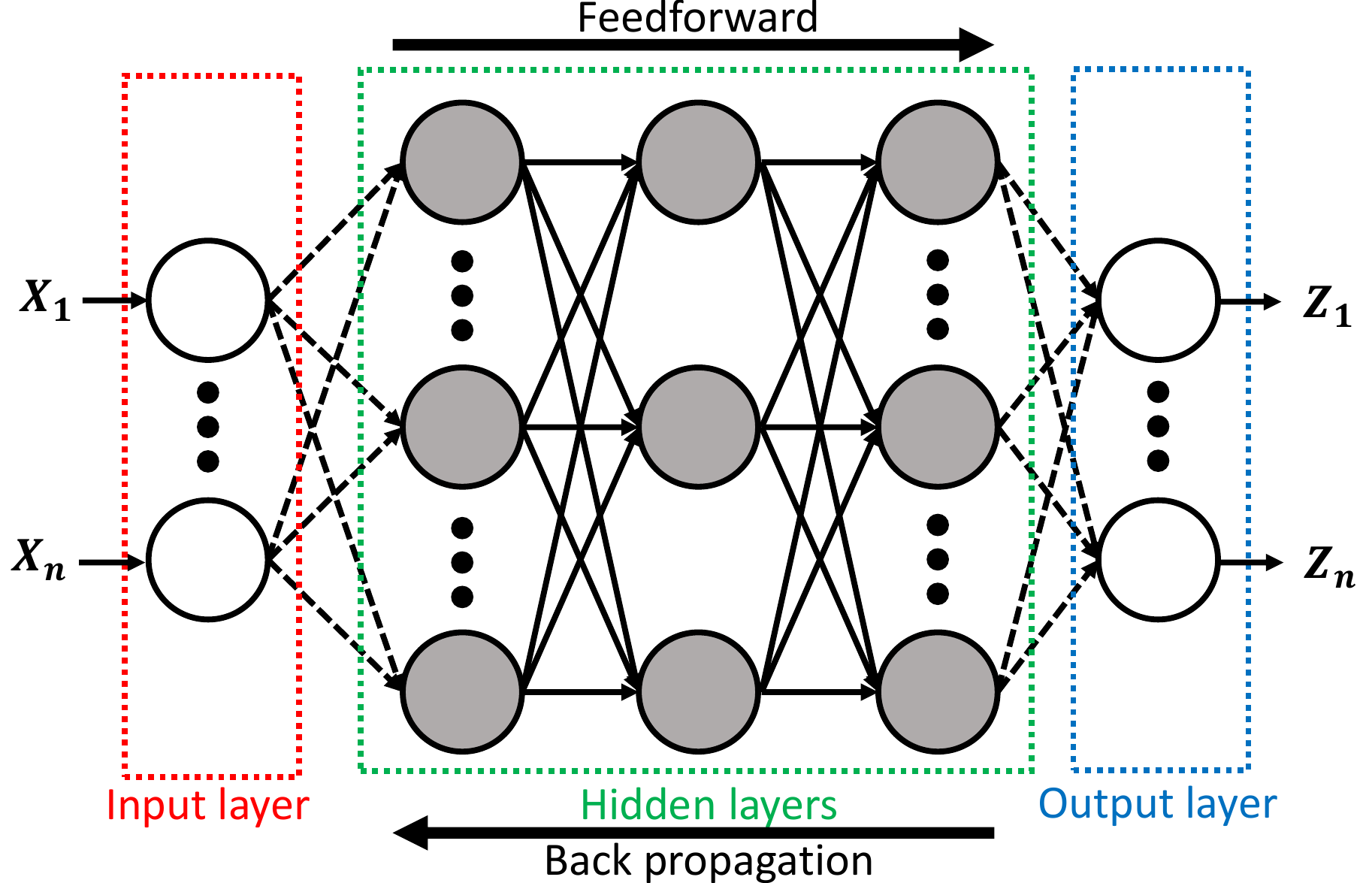}
    \caption{Typical architecture of a fully connected neural network.}
    \label{DNN}
\end{figure}
The architecture of a DNN can be characterized by the number of hidden layers and the number of neurons in each hidden layer. The neurons of consecutive layers are connected by a set of weights $\bm{W}$ and biases $\bm{b}$. The output $\bm{y}^i$ of layer $i$ can be related to its corresponding weights and biases as:
\begin{equation}
    \bm{y}^i = f_{\rm{act}}^i( \bm{W}^i \bm{y}^{i-1} + \bm{b}^i ),
\end{equation}
where $f_{\rm{act}}^i$ denote the activation function for this layer. Training of the DNN refers to the iterative process where the weights and biases of the model are updated by means of gradient descent \cite{pattanayak2017pro} such that the value of a user-defined loss function is minimized.

\subsection{The deep energy method}
\label{DEM}
In this section, we describe DEM in the context of the elasticity equation in structural mechanics. We consider a homogeneous, isotropic, linear elastic body undergoing small deformation. We remark that DEM is not limited to linear elastic materials and has been applied in nonlinear hyperelastic material models \cite{nguyen2020deep}. In the absence of any body and inertial forces, the equilibrium equation in tensorial form reads:
\begin{equation}
\begin{aligned}
    \nabla \cdot \bm{\sigma} = \bm{0}, \;\; \forall \bm{X} \in \Omega.
\end{aligned}
\end{equation}
The system is subject to Dirichlet boundary condition:
\begin{equation}
\begin{aligned}
    \bm{ u } = \Bar{\bm{u}}, \;\; \forall \bm{X} \in \partial \Omega_u,\\
\end{aligned}
\end{equation}
and/or Neumann boundary condition:
\begin{equation}
\begin{aligned}
    \bm{\sigma} \cdot \bm{ n } = \Bar{\bm{t}} , \;\; \forall \bm{X} \in \partial \Omega_t.
\end{aligned}
\end{equation}
In the small deformation setting, the strain tensor is given by:
\begin{equation}
    \bm{\epsilon} = \frac{1}{2} ( \nabla \bm{u} + \nabla \bm{u}^T ).
    \label{strain}
\end{equation}
Stress can be computed from the constitutive law as:
\begin{equation}
    \bm{\sigma} = \frac{E}{1+\nu} \bm{\epsilon} + \frac{E\nu}{ (1+\nu)(1-2\nu) } tr(\bm{\epsilon}) \bm{I},
    \label{stress}
\end{equation}
where $E$ and $\nu$ are the Young's modulus and Poisson's ratio. DEM seeks the solution to the equilibrium equation via the principle of minimum potential energy. For a body in static equilibrium with no applied body forces, the potential energy of the system reads:
\begin{equation}
    \psi(\bm{u}) = \frac{1}{2} \int_{\Omega} \bm{\sigma} : \bm{\epsilon} \, dV - \int_{\partial \Omega_t} \Bar{\bm{t}} \cdot \bm{u} \, dA.
    \label{PE}
\end{equation}
The loss function ($\mathcal{L}$) in DEM is defined identically as the potential of the system:
\begin{equation}
    \mathcal{L}( \bm{u} ) = \psi(\bm{u}).
    \label{loss}
\end{equation}
The training of the DEM model $\mathcal{M}: \mathbb{R}^n \to \mathbb{R}^n / \bm{X} \Rightarrow \Tilde{\bm{u}}(\bm{X}) $ \footnote{Note that the immediate output of DEM model is $\Tilde{\bm{u}}$, which is different from $\bm{u}$ in \eref{strain} and \eref{PE}. } can be viewed as minimizing the loss function (system potential energy) $\mathcal{L}$ through an optimizer. The solution to the elasticity problem, as given by the DEM model, is defined as:
\begin{equation}
    \bm{u}^* = {\rm{arg}}\min_{ \bm{u} } \, \mathcal{L}( \bm{u} ).
\end{equation}
In this work, the L-BFGS optimizer \cite{zhu1997algorithm} was used to train the model. The learning rate is another parameter that can be changed by the user to achieve optimal framework performance. Similar to the early stopping schedule commonly used in model training, we monitor the relative change of the loss function and stop the training once the relative change in loss function value is less than the user-specified tolerance $\epsilon_{tol}$. 

The underlying architecture of the DEM model is a fully connected feed-forward network, which is similar to the one depicted in \fref{DNN} and was described in the work of Nguyen et al. \cite{nguyen2020deep}. Fourier transform was applied using the package Random Fourier Features Pytorch (RFF) \cite{long2021rffpytorch} to the input features to transform them to frequency domain, which was shown to increase the accuracy of PINNs in the work of Wang et al. \cite{wang2021eigenvector}. The NN architecture is fully parametrizable, and is characterized by the number of layers, number of neurons (per layer), the activation function, and two standard deviation values used to initialize the weights and biases of the MLP and RFF. Hyperparameter optimization can be performed to improve the performance of the DEM model.

\subsection{Enforcement of boundary conditions}
In the context of elasticity problems, three different types of boundary conditions (BCs) are common, namely the Neumann, Dirichlet, and periodic BCs. Enforcement of Neumann BCs is straightforward. The value of the prescribed traction $\Bar{\bm{t}}$ enters directly in the boundary integral part of \eref{PE}, while the boundary displacements are provided by the DEM model. Traction-free surfaces are satisfied trivially as the boundary integral vanishes on these surfaces. A surface (in 3D) or edge (in 2D) integration rule needs to be defined for numerically evaluating the surface integral in regions where traction is prescribed. 

Dirichlet BCs are typically enforced via the penalty method by adding a penalty term to the loss function \cite{nguyen2020deep,abueidda2022deep,abueidda2021meshless}:
\begin{equation}
    MSE_{\partial \Omega_u } = \frac{1}{N_u} \sum_{j=1}^{N_u} || \bm{u} - \Bar{\bm{u}} ||^2,
\end{equation}
leading to a modified loss function definition:
\begin{equation}
    \mathcal{L}( \bm{u} ) = \psi(\bm{u}) + w \, MSE_{\partial \Omega_u },
\end{equation}
where $w$ is a user-defined weight parameter. To avoid defining the additional parameter $w$, we adopted a direct approach to enforce Dirichlet BCs via an additive decomposition:
\begin{equation}
    \bm{u}(\bm{X}) = \Tilde{\bm{u}}(\bm{X}) * \bm{m}(\bm{X}) + \bm{u}_0(\bm{X}),
\end{equation}
where $*$ denotes element-wise multiplication between two vectors, the vectors $\bm{m}$ and $\bm{u}_0$ are chosen such that:
\begin{equation}
    \bm{m}(\bm{X}) = \bm{0} \,\, {\rm{and}} \,\, \bm{u}_0 (\bm{X}) = \Bar{\bm{u}}, \forall \bm{X} \in \partial \Omega_u
\end{equation}
By design, $\bm{u}(\bm{X})$ satisfies all Dirichlet boundary conditions for all arbitrary $\Tilde{\bm{u}}$ produced by the DEM model.

Periodic BCs are commonly used when simulating a representative volume element that repeats infinitely in space. It is also used when simulating unit cells of periodic meta materials. In the context of the FEM, periodic BCs are typically stated in the form of constraint equations \cite{xia2015design}:
\begin{equation}
    u_i^{k+} - u_i^{k-} = \epsilon^0_{ij} ( x_j^{k+} - x_j^{k-} ),
    \label{PBC}
\end{equation}
where $k-$ and $k+$ denote the left and right instances of the $k^{th}$ periodic boundary pair, and $\epsilon^0_{ij}$ is the (known) applied strain tensor. For easier convergence of the DEM training process, we adopted a mixed implementation for periodic BCs, where we treat zero and non-zero applied strain components differently. The boundary displacement induced by a non-zero applied strain component is enforced as a pair of Dirichlet BCs, and zero applied strain components are enforced by penalty. Without loss of generality, let the non-zero component of $\epsilon^0_{ij}$ be $\epsilon_a$, which is applied on the $k^{th}$ periodic boundary pair. This leads to two Dirichlet boundary conditions:
\begin{equation}
\begin{aligned}
    \bm{u}(\bm{X}) = \bm{0}, \;\; \forall \bm{X} \in \partial \Omega_{k-},\\
    \bm{u}(\bm{X})_a = \Delta L_k \epsilon_a , \;\; \forall \bm{X} \in \partial \Omega_{k+},
\end{aligned}
\end{equation}
where $\Delta L_k$ is the distance between the $k+$ and $k-$ boundary pair. For all other boundary pairs that correspond to zero applied strain components, the constraint equation \eref{PBC} reduces to:
\begin{equation}
    u_i^{k+} = u_i^{k-}.
    \label{PBC_zero}
\end{equation}
Therefore, the following penalty term is added to the loss function:
\begin{equation}
    MSE_{PBC } = \frac{E}{N_p} \sum_{j=1}^{N_p} || u_i^{k+} - u_i^{k-} ||^2,
\end{equation}
where $E$ is the Young's modulus of the material (a physics-informed penalty weight) and $N_p$ is the number of points of the $k^{th}$ periodic boundary.

\subsection{Topology optimization}
In this section, we describe topology optimization (TO) in the context of DEM. The mathematical formulation of the TO problem using DEM can be stated as:
\begin{equation}
\begin{aligned}
    \min_{ \bm{\rho} } f( \bm{u} , \bm{\rho} )\\
    {\rm{s.t.}} \; \bm{u} = {\rm{arg}}\min_{ \bm{u} } \, \mathcal{L}( \bm{u} , \bm{\rho} ), \\
    V_e \sum \bm{\rho} = \Bar{V}, \\
    0 \le \rho_e \le 1, \, \forall e = 1 \cdots N,
\end{aligned}
\end{equation}
where $f( \bm{u} , \bm{\rho} )$, $\bm{\rho}$, $V_e$ and $\Bar{V}$ denote the objective function to minimize, element density vector, element volume and target volume, respectively. A typical objective of TO is to minimize the system compliance subject to a volume/mass constraint. For compliance minimization, the objective function is:
\begin{equation}
    f_{ \rm{comp} } ( \bm{u} , \bm{\rho} ) = \frac{1}{2} \int_\Omega  \bm{\sigma}^*(\bm{\rho}) : \bm{\epsilon} \, dV
    \label{min_comp}
\end{equation}
where $\bm{\sigma}^*(\bm{\rho})$ is the stress computed from the classical solid isotropic material penalization (SIMP) scheme \cite{rozvany2009critical}:
\begin{equation}
    \bm{\sigma}^*(\bm{\rho}) = \bm{\rho}^p * \bm{\sigma}.
    \label{SIMP_stress}
\end{equation}
The penalty exponent $p$ is taken to be 3 in this work.

For designing periodic 2D meta structures, a commonly employed objective is to maximize the homogenized shear modulus of the unit cell \cite{kollmann2020deep,xia2015design,zhang2019topology}:
\begin{equation}
    G_{\rm{homo}} = \frac{1}{A} \int_\Omega  \bm{\sigma}^{p*}(\bm{\rho}) : \bm{\epsilon}^{p} \, dA,
    \label{shear_mod}
\end{equation}
where $A$ is the area of the 2D element, $\bm{\sigma}^{p}$ and $\bm{\epsilon}^{p}$ denote the stress and strain fields computed from the \emph{periodic} displacement field subjected to the following applied simple shear loading ($\bm{\epsilon}^0$ in \eref{PBC} ):
\begin{equation}
    \bm{\epsilon}^0 = \begin{bmatrix}
    0 & 0.01 \\
    0.01 & 0
    \end{bmatrix}.
\end{equation}
To fit into a minimization framework, the objective function is defined as:
\begin{equation}
    f_{ \rm{shear} } ( \bm{u} , \bm{\rho} ) = -G_{\rm{homo}}.
    \label{max_shear}
\end{equation}

Due to the introduction of the element density vector, we correspondingly modify the definition of the potential energy and loss function as:
\begin{equation}
    \psi( \bm{u} , \bm{\rho} ) = \mathcal{L}( \bm{u} , \bm{\rho} ) = \frac{1}{2} \int_{\Omega}  \bm{\sigma}^*(\bm{\rho}) : \bm{\epsilon} \, dV - \int_{\partial \Omega_t} \Bar{\bm{t}} \cdot \bm{u} \, dA.
    \label{PE_SIMP}
\end{equation}

In this work, we chose to evaluate the volume integral in \eref{PE_SIMP} by Gauss quadrature integration. This necessitates the formation of quadrilateral (in 2D) or hexahedral (in 3D) isoparametric finite elements from the structured nodes $\bm{X}$ in the DEM domain, which is very similar to a FEM-type treatment. The need to form elements is a major disadvantage of the framework, as compared to other meshless DEM methods like those developed in \cite{abueidda2021meshless,nguyen2020deep} which only require nodes but not elements to exist in the simulation domain. The gradient operator in \eref{strain} is evaluated by the gradients of the finite element shape functions, and not from automatic differentiation of the DEM model. The density vector $\bm{\rho}$ is defined at the cell center, identical in classical FEM-based TO. The method of moving asymptotes (MMA) \cite{svanberg1987method} is used to iteratively update the density based on the sensitivity.

\subsection{Sensitivity analysis}
In this section, we describe the sensitivity analysis in the context of DEM. Comparing \eref{min_comp} and \eref{max_shear}, we see that both are similar in form to a classical compliance minimization problem in FEM-based TO. Since the compliance minimization problem is self-adjoint, the adjoint vector is simply the negative of the displacement, and therefore, no additional adjoint analysis is needed to evaluate the sensitivity. In this case, one simply needs to express the sensitivity in terms of the displacement field and its gradients produced by DEM. For compliance minimization problem (\eref{min_comp}), the sensitivity of element $e$ is given by:
\begin{equation}
    \frac{ \partial f_{ \rm{comp} } }{ \partial \rho_e } = -\frac{1}{2} p \rho_e^{p-1} \int_{\Omega_e} \bm{\sigma} : \bm{\epsilon} \, dV.
\end{equation}
Similarly, for maximization of shear modulus in 2D meta material unit cells (\eref{max_shear}), the sensitivity of element $e$ is given by:
\begin{equation}
    \frac{ \partial f_{ \rm{shear} } }{ \partial \rho_e } = \frac{1}{A_e} p \rho_e^{p-1} \int_{\Omega_e} \bm{\sigma} : \bm{\epsilon} \, dA.
\end{equation}

To avoid checkerboarding and mesh-dependence, the density filtering technique introduced by Bruns et al. \cite{bruns2001topology} was employed. In this approach, the MMA solver does not directly manipulate $\bm{\rho}$, but instead manipulates on a pseudo-density (the design variables) $\bm{\xi}$ that is related to $\bm{\rho}$ as:
\begin{equation}
    \bm{\rho} = \bm{W} \bm{\xi},
\end{equation}
where $\bm{W}$ is the normalized filter matrix whose rows are given by the normalized weights $\Bar{q}_{ij}$:
\begin{equation}
\begin{aligned}
    q_{ij} = {\rm{max}}( r_{min} - || \bm{X}_i - \bm{X}_j || ), \\
    \Bar{q}_{ij} = \frac{1}{\sum_{k=1}^N q_{ik} } q_{ij}.
\end{aligned}
\end{equation}
Algorithm \ref{DEM_TO} summarizes the process of TO using the DEM method.

\begin{algorithm}[!ht]
\DontPrintSemicolon
    \KwInput{ Network architecture, domain size, grid size, material properties, $\epsilon_{tol}$ , $r_{min}$, $\Bar{V}$, initial density distribution $\bm{\xi}^0$, maximum TO iterations}
    
    \KwOutput{Optimized design defined by density array $\bm{\rho}$ }

    \tcc{Initialization}
    Build filter matrix $\bm{W}$\\
    Initialize weights and biases of the DEM model $\mathcal{M}$ \\
    $ i \gets 0 $
    
    \tcc{Begin topology optimization}
    \While{ $i < $ max TO iteration }{
    Apply filter: $\bm{\rho}^i = \bm{W} \bm{\xi}^i$
    
    \tcc{Begin DEM training}
    \While{ not converged }{
        Obtain $\Tilde{\bm{u}}$ from $\mathcal{M}$, apply Dirichlet BCs to get $\bm{u}$ \\
        Use shape function gradient operator to compute $\bm{\epsilon}$ \\
        Use $\bm{\rho}$ to compute $\bm{\sigma}^*$ \\
        Compute loss \\
        Update the weights and biases of $\mathcal{M}$ through back-propagation
    }
    
    \tcc{Begin sensitivity calculation}
    Obtain converged $\Tilde{\bm{u}}$ from $\mathcal{M}$, apply Dirichlet BCs\\
    Compute $\bm{\epsilon}$ and $\bm{\sigma}^*$ \\
    Compute the element-wise filtered sensitivity \\
    Invert filtering for sensitivity: $\frac{\partial f }{ \partial \bm{\xi}} = \bm{W}^T \frac{\partial f }{ \partial \bm{\rho}} $ \\
    Pass sensitivity to MMA solver to obtain $\bm{\xi}^{i+1}$ \\
    $ i \gets i + 1 $
    }

\caption{Topology optimization using the deep energy method}
\label{DEM_TO}
\end{algorithm}

\section{Results and discussion}
\label{sec:results}
In this section, we present three numerical examples that showcase the capabilities of our DEM-based TO framework. In the first example, we performed compliance minimization in 2D for two different geometries: 1. a long slender bridge fixed on both ends subject to center downward load, and 2. a short cantilever beam subject to downward load on its right free end. In the second example, we extend the bridge example to 3D. In the last example, we maximize the shear modulus of 2D periodic unit cells to design meta materials, and compare the results at different initial conditions. In all cases, we compared our DEM results with those obtained from FEM. All training of the DEM model was done on Delta, an HPC cluster hosted at the National Center for Supercomputing Applications (NCSA). A representative GPU compute node on Delta has the following hardware specifications: one AMD Milan CPU, four NVIDIA A40 GPUs, and 256 GB RAM. The hyperparameters of the DEM model are provided in \tref{hyper}, and were obtained following the procedure described in the work by Chadha et al. \cite{chadha2022optimizing}.
\begin{table}[h!]
    \caption{Hyper parameters of the DEM model}
    \small
    \centering
    \begin{tabular}{ccccccccc}
      & \vline & Layers  & Neurons & Activation function & Max iteration & learning rate & $\sigma_{MLP}$ & $\sigma_{RFF}$ \\
    \hline
    Value & \vline & 5 & 68 & RReLU & 100 & 1.735 & 0.0622 & 0.1192
    \end{tabular}
    \label{hyper}
\end{table}
The DEM model was implemented in PyTorch (version 1.11.0) \cite{NEURIPS2019_9015}. Material properties used in all three examples are: $E=200$ MPa, and $\nu = 0.3$. Plane stress condition and unit thickness were assumed for all 2D simulations. A target volume fraction of 40\% was used. A Python implementation of the MMA solver by Chandrasekhar et al. \cite{chandrasekhar2021auto} was used. For all DEM examples presented, a total of 80 TO iterations were conducted using a filter radius of 0.25 m.

\subsection{Compliance minimization in 2D}
\label{2d_comp}
Two different cases are presented in this example. In the first case, a 2D rectangular bridge of size 12-by-2 m$^2$ was subjected to a downward surface traction at the center of the top edge over a length of 0.5 m. The bridge was fixed at its left and right edges and 3751 nodes were placed into the domain forming a 121-by-31 grid. In the second case, a 2D rectangular beam of size 10-by-5 m$^2$ was subjected to a downward concentrated load at the center of the right edge, while the left edge was fixed. 4186 nodes were placed into the domain, forming a 91-by-46 grid. The relative convergence tolerance $\epsilon_{tol}$ was set to $5\times10^{-6}$ for all DEM training. As comparison, FEM-based TO was applied to solve the same problem using identical node layout. 

\begin{figure}[h!]
    \newcommand\x{0.23}
    \centering
    \begin{tabular}{ c  c  c  c  }
    \begin{minipage}[c]{\x\textwidth}
       \centering 
        \subfloat[DEM, iteration 2]{\includegraphics[trim={0.5cm 6.5cm 2.cm 1.5cm},clip,width=\textwidth]{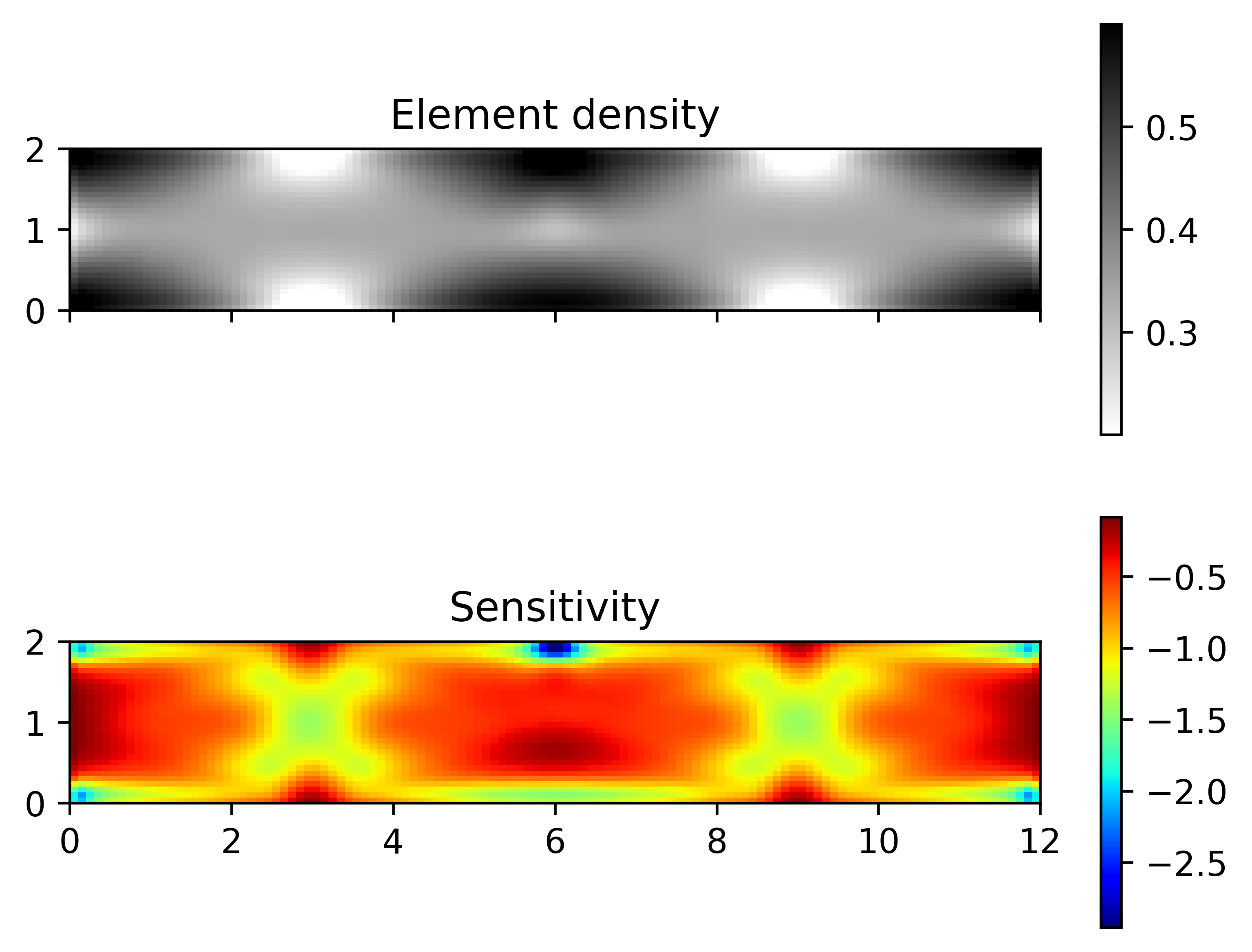}
        \label{fig:d1}}
    \end{minipage}
    &
    \begin{minipage}[c]{\x\textwidth}
       \centering 
        \subfloat[DEM, iteration 15]{\includegraphics[trim={0.5cm 6.5cm 2.cm 1.5cm},clip,width=\textwidth]{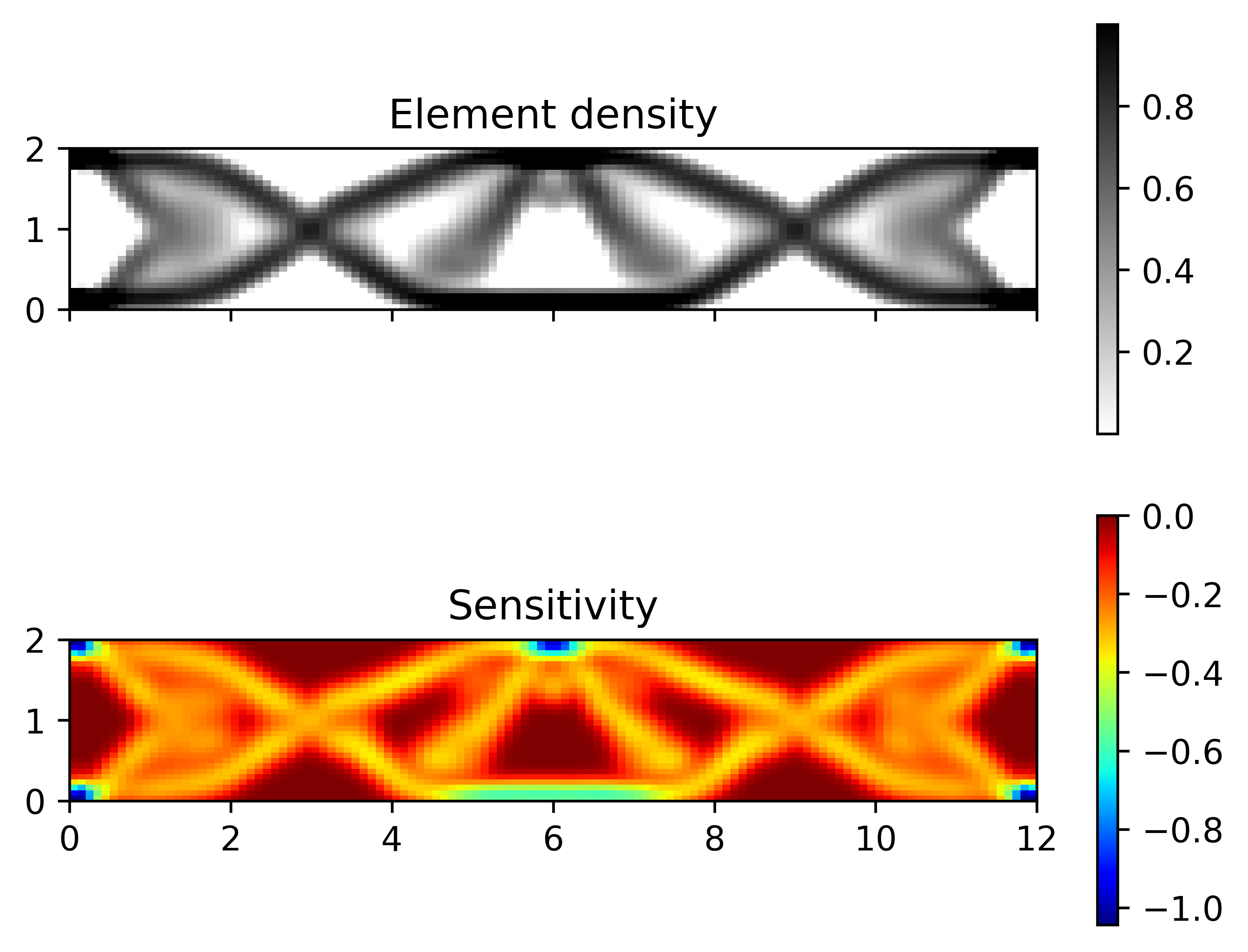}
        \label{fig:d2}}
    \end{minipage}
    &
    \begin{minipage}[c]{\x\textwidth}
       \centering 
        \subfloat[DEM, iteration 30]{\includegraphics[trim={0.5cm 6.5cm 2.cm 1.5cm},clip,width=\textwidth]{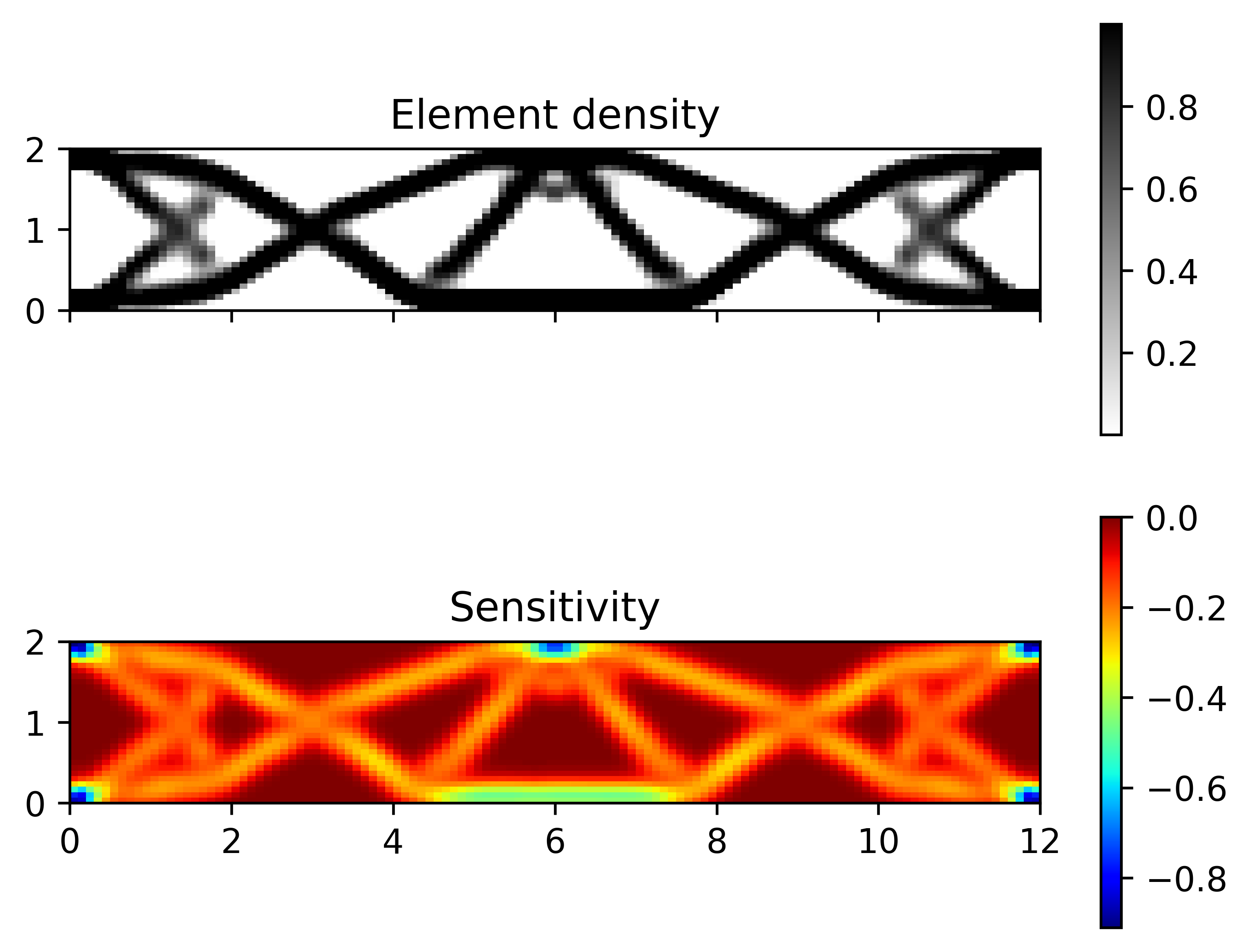}
        \label{fig:d3}}
    \end{minipage}
    &
    \begin{minipage}[c]{\x\textwidth}
       \centering 
        \subfloat[DEM, iteration 80]{\includegraphics[trim={0.5cm 6.5cm 2.cm 1.5cm},clip,width=\textwidth]{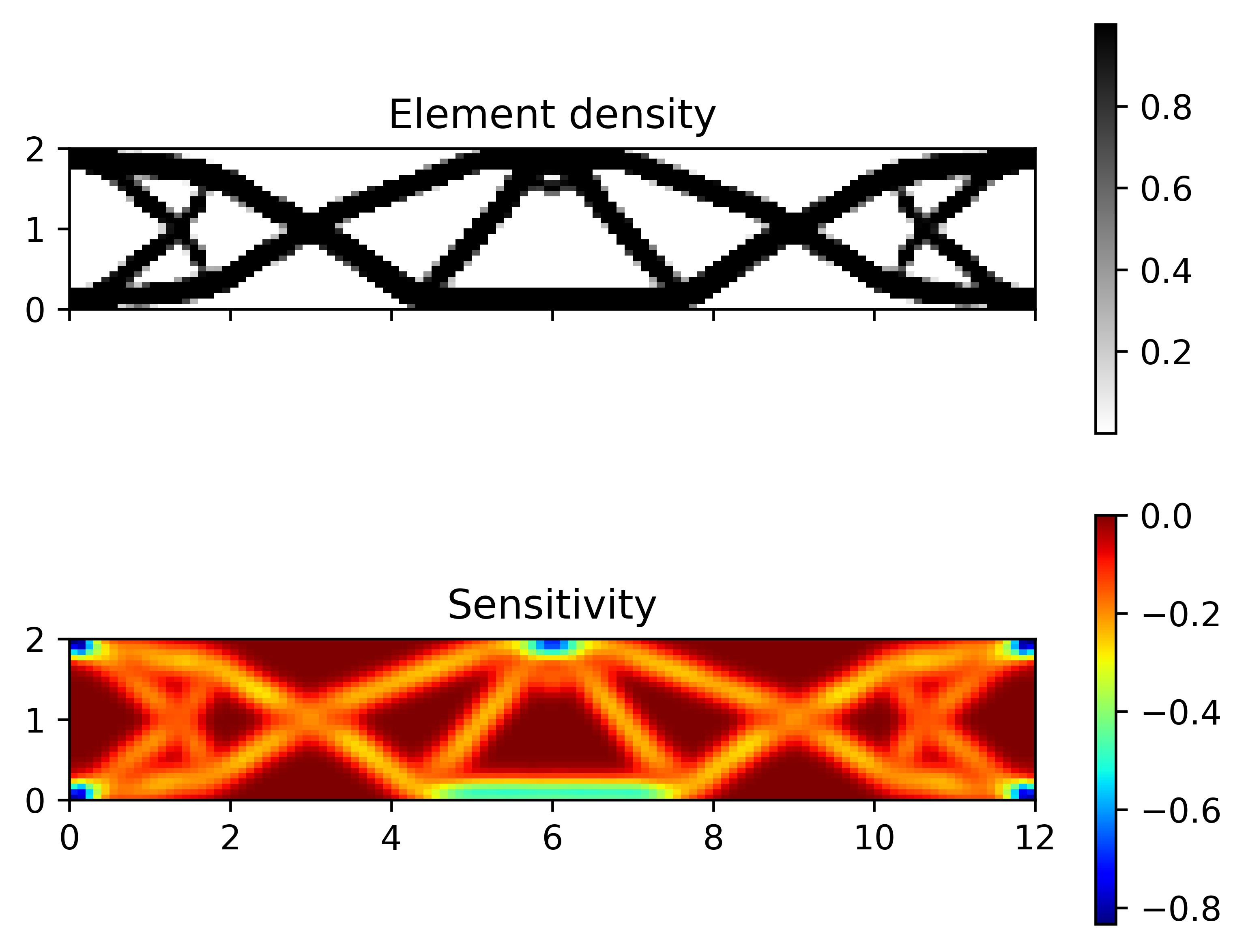}
        \label{fig:d4}}
    \end{minipage}\\
    
    & \\
    
    \begin{minipage}[c]{\x\textwidth}
       \centering 
        \subfloat[FEM, iteration 2]{\includegraphics[trim={0.5cm 6.5cm 1.9cm 1.4cm},clip,width=\textwidth]{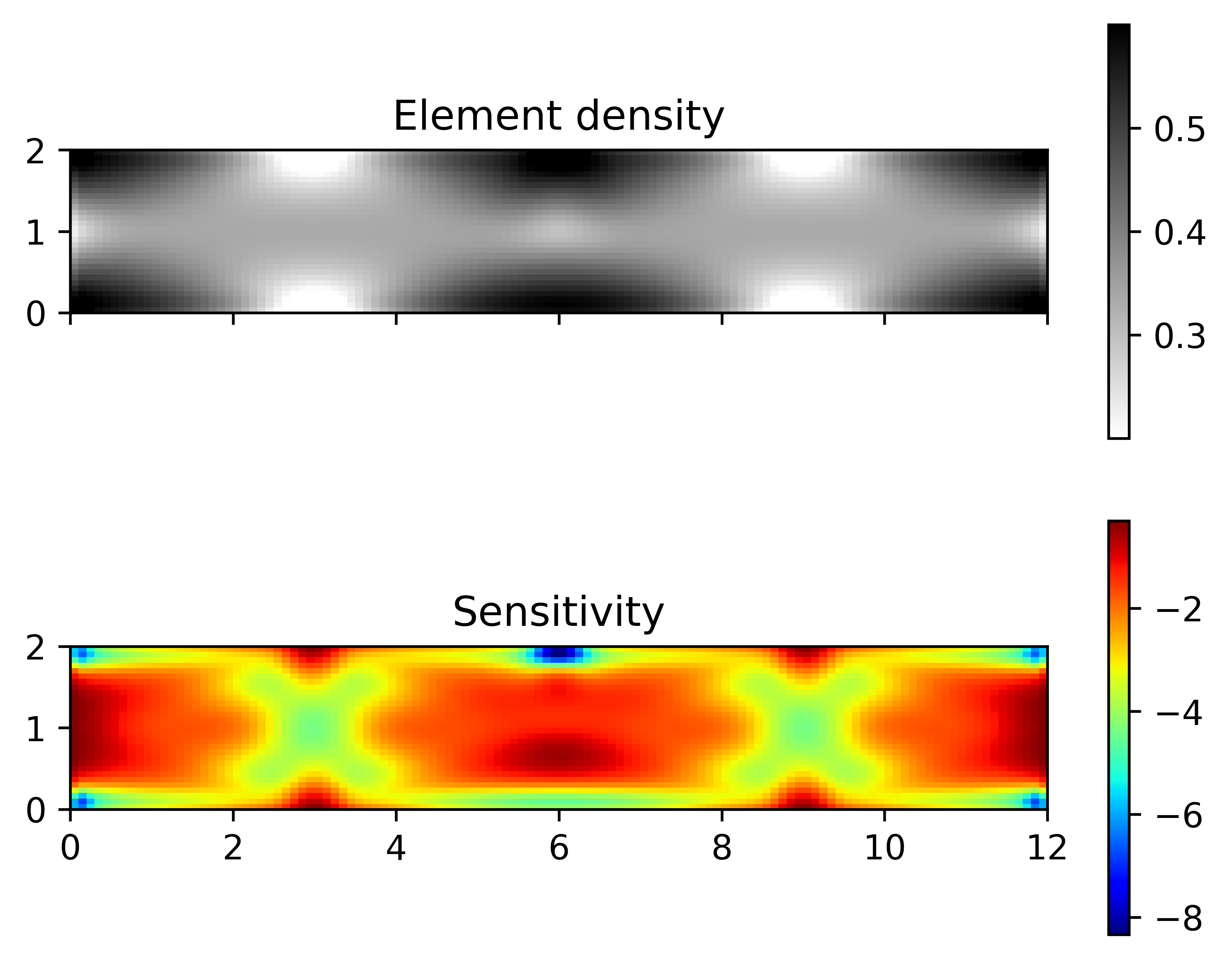}
        \label{fig:d5}}
    \end{minipage}
    &
    \begin{minipage}[c]{\x\textwidth}
       \centering 
        \subfloat[FEM, iteration 15]{\includegraphics[trim={0.5cm 6.5cm 1.9cm 1.4cm},clip,width=\textwidth]{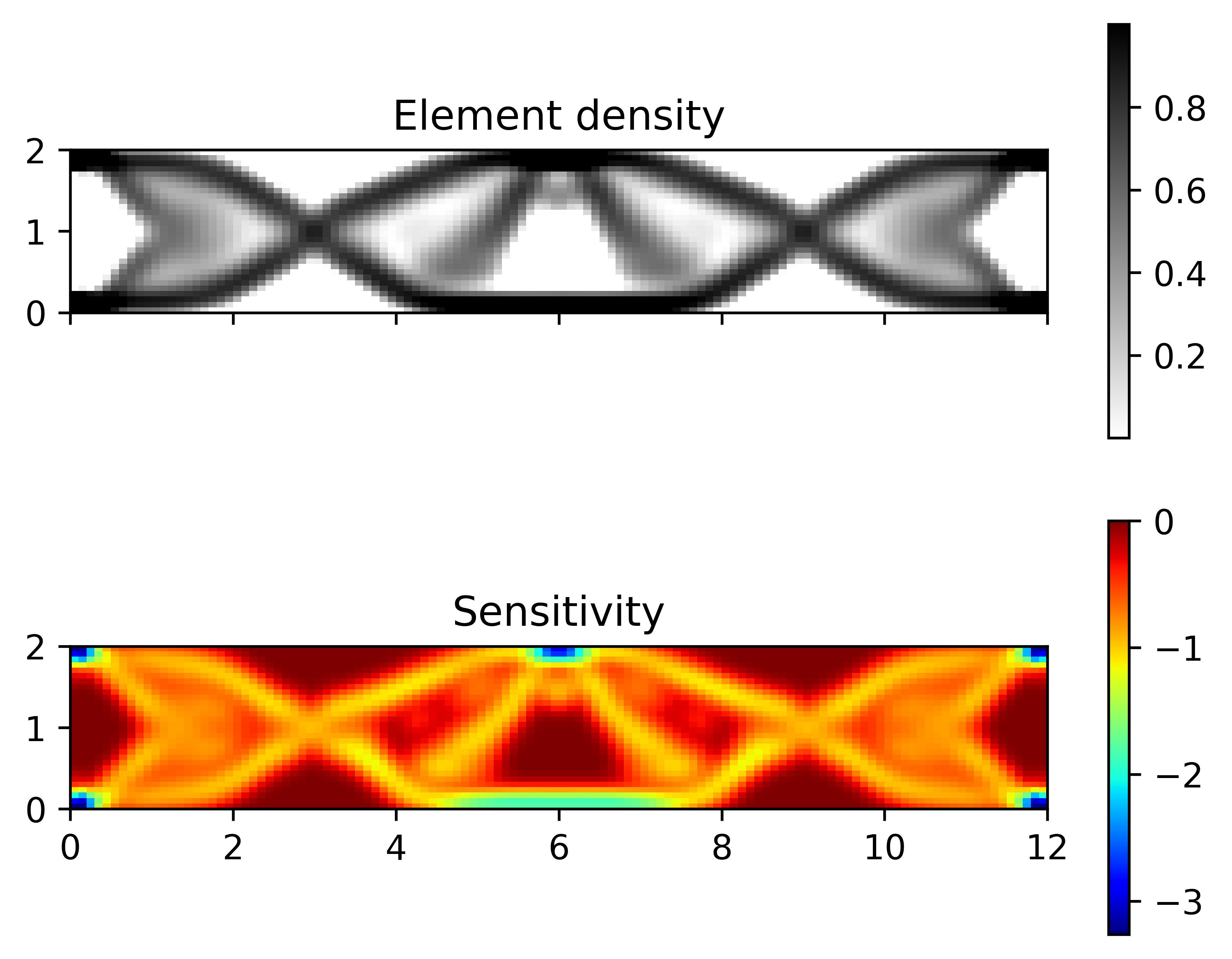}
        \label{fig:d6}}
    \end{minipage}
    &
    \begin{minipage}[c]{\x\textwidth}
       \centering 
        \subfloat[FEM, iteration 30]{\includegraphics[trim={0.5cm 6.5cm 1.9cm 1.4cm},clip,width=\textwidth]{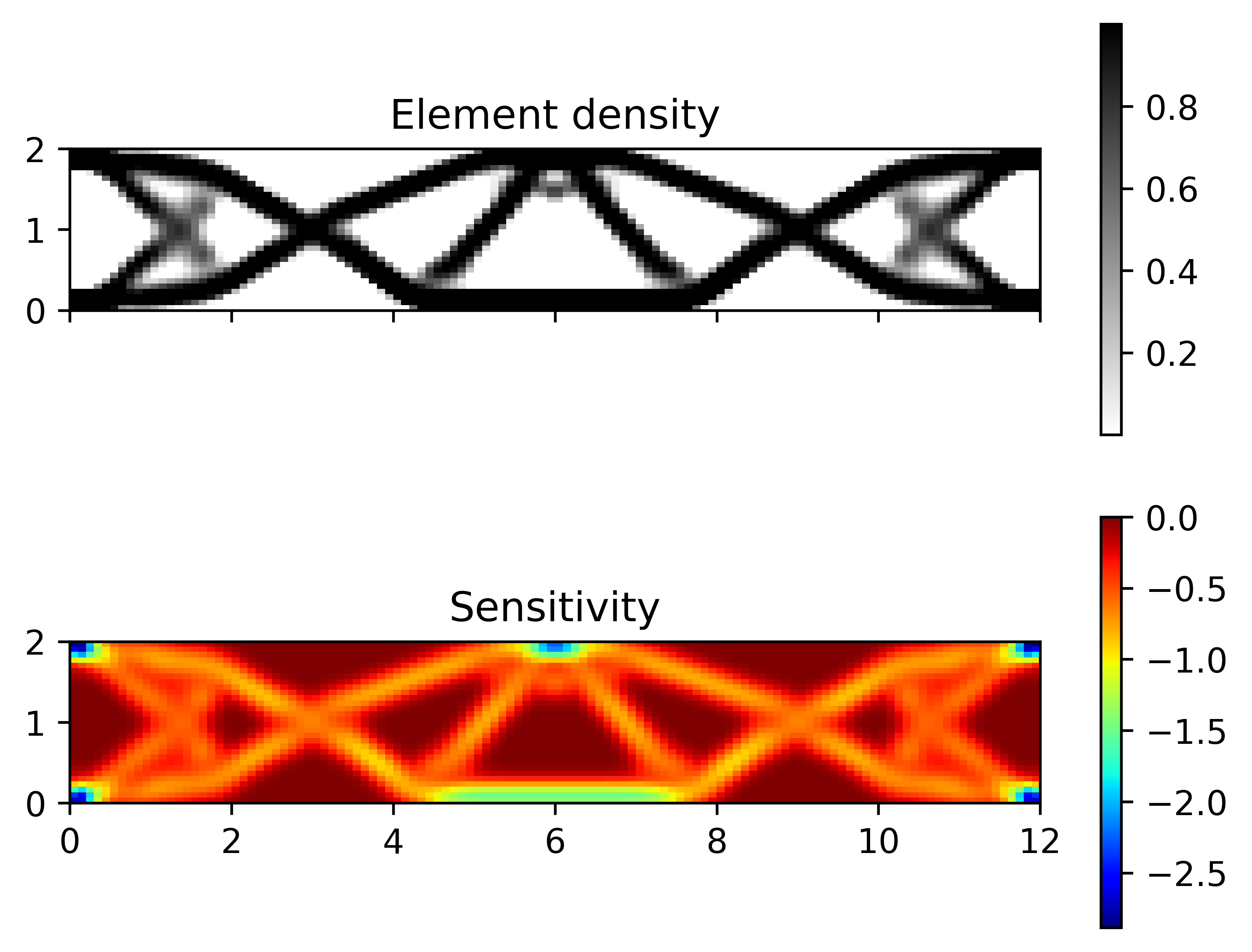}
        \label{fig:d7}}
    \end{minipage}
    &
    \begin{minipage}[c]{\x\textwidth}
       \centering 
        \subfloat[FEM, iteration 80]{\includegraphics[trim={0.5cm 6.5cm 1.9cm 1.4cm},clip,width=\textwidth]{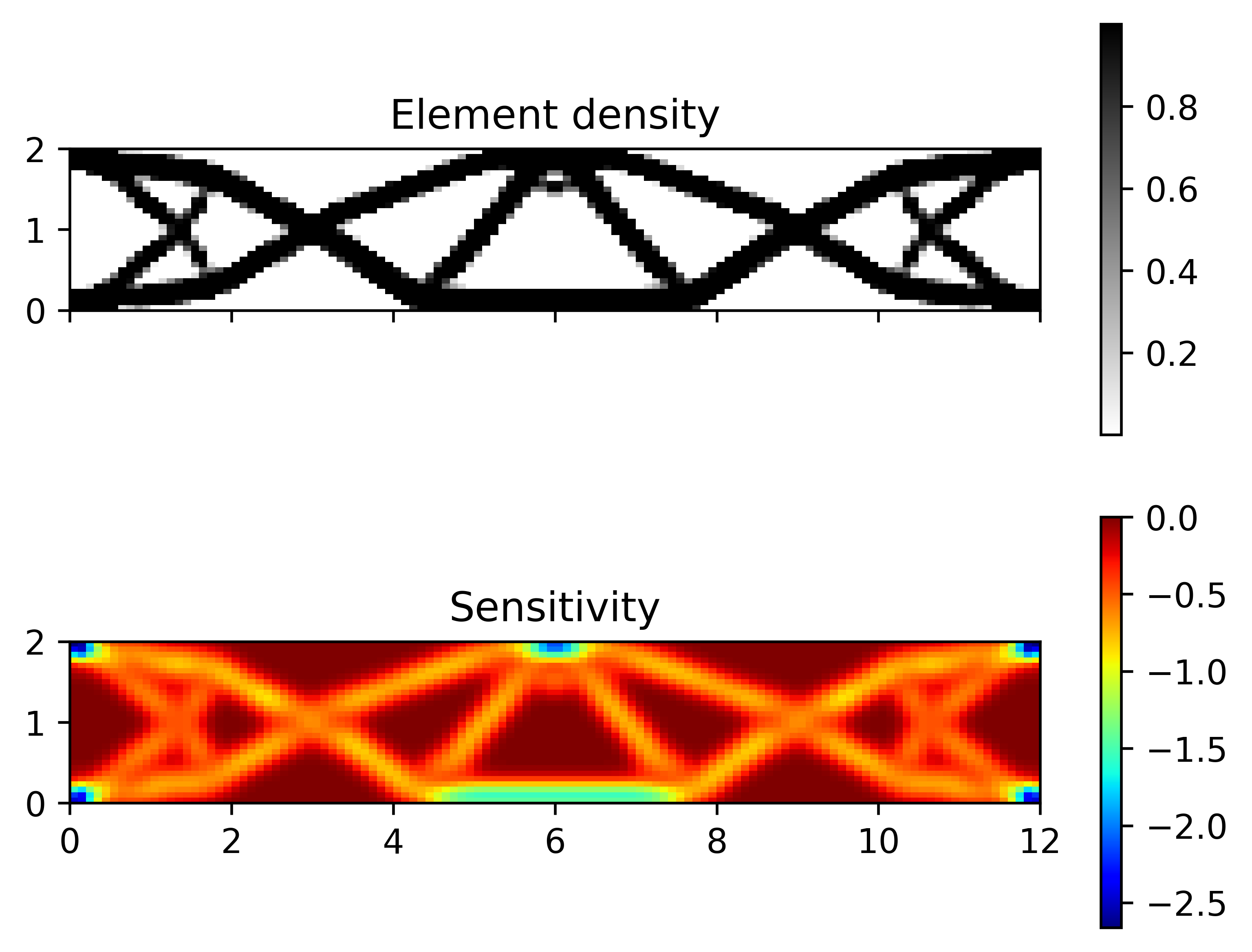}
        \label{fig:d8}}
    \end{minipage}

    \end{tabular}
    \caption{Density plot of designs generated by DEM and FEM for the bridge example. Both left and right edges are fixed, and a downward load is applied on the center of the top edge.}
    \label{case1_FEM-DEM}
\end{figure}

\begin{figure}[h!]
    \newcommand\x{0.23}
    \centering
    \begin{tabular}{ c  c  c  c  }
    \begin{minipage}[c]{\x\textwidth}
       \centering                                         
        \subfloat[DEM, iteration 2]{\includegraphics[trim={0.5cm 5.8cm 2.3cm 0.65cm},clip,width=\textwidth]{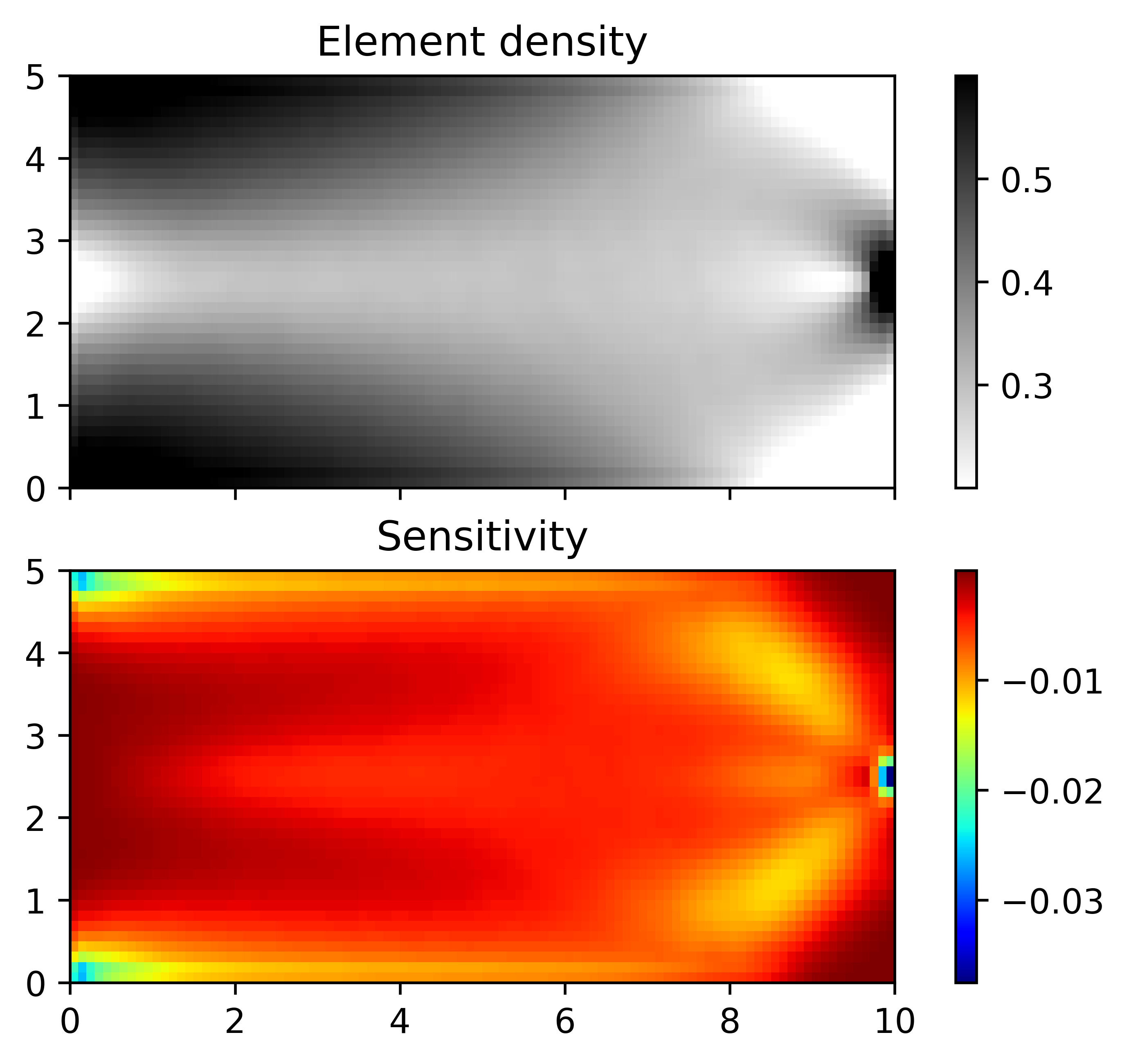}
        \label{fig:d1}}
    \end{minipage}
    &
    \begin{minipage}[c]{\x\textwidth}
       \centering 
        \subfloat[DEM, iteration 15]{\includegraphics[trim={0.5cm 5.8cm 2.3cm 0.65cm},clip,width=\textwidth]{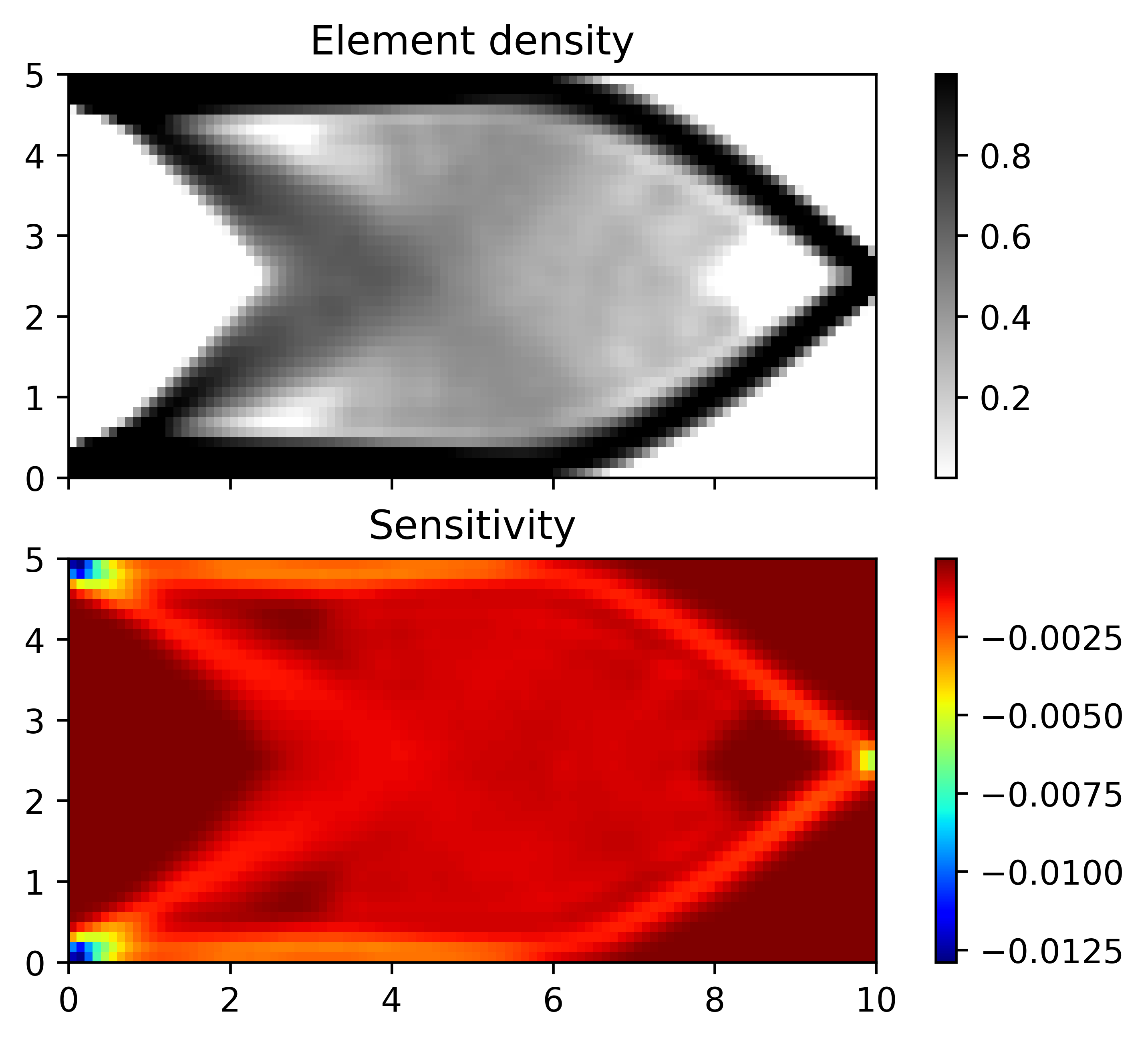}
        \label{fig:d2}}
    \end{minipage}
    &
    \begin{minipage}[c]{\x\textwidth}
       \centering 
        \subfloat[DEM, iteration 30]{\includegraphics[trim={0.5cm 5.8cm 2.3cm 0.65cm},clip,width=\textwidth]{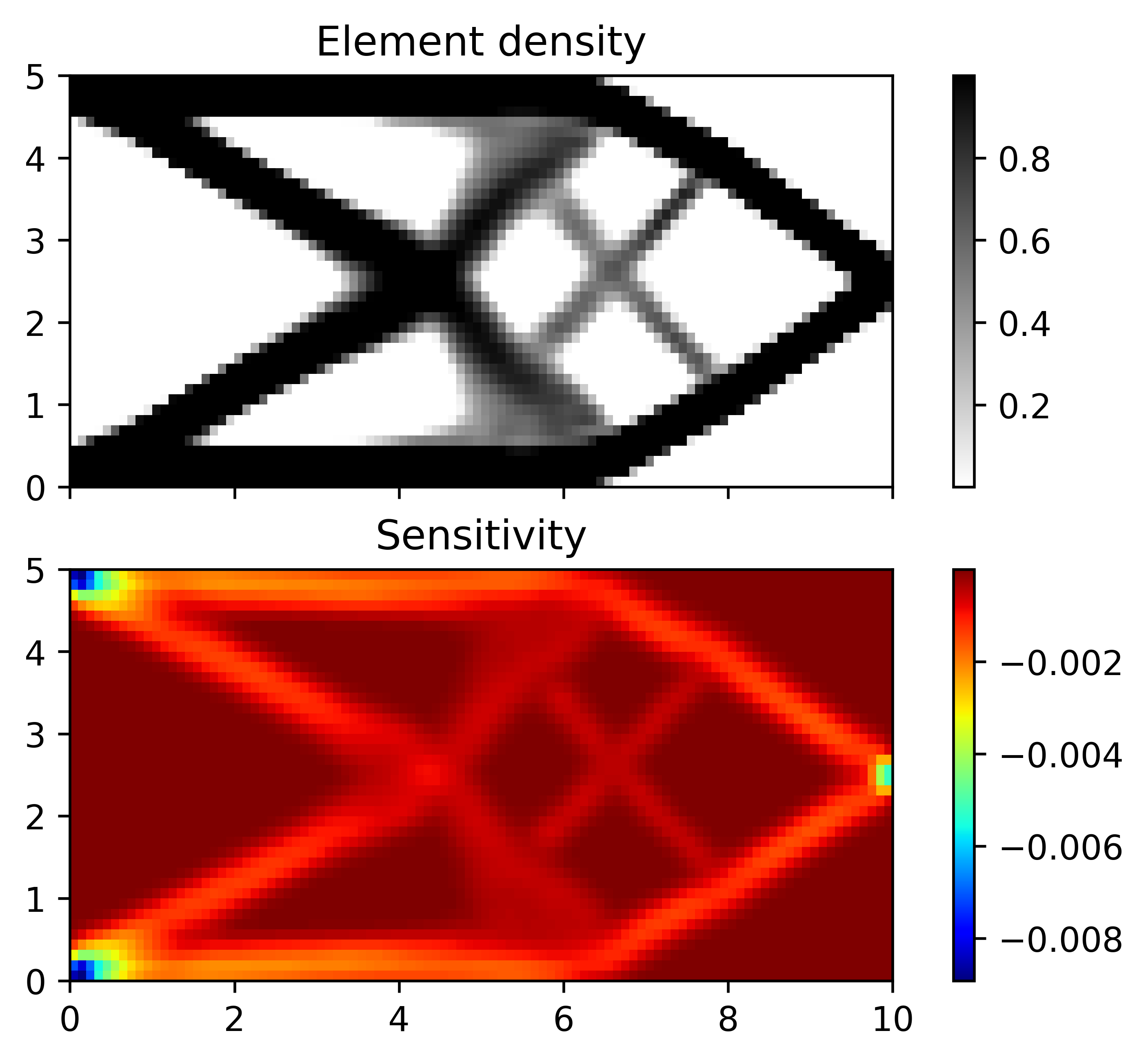}
        \label{fig:d3}}
    \end{minipage}
    &
    \begin{minipage}[c]{\x\textwidth}
       \centering 
        \subfloat[DEM, iteration 80]{\includegraphics[trim={0.5cm 5.8cm 2.3cm 0.65cm},clip,width=\textwidth]{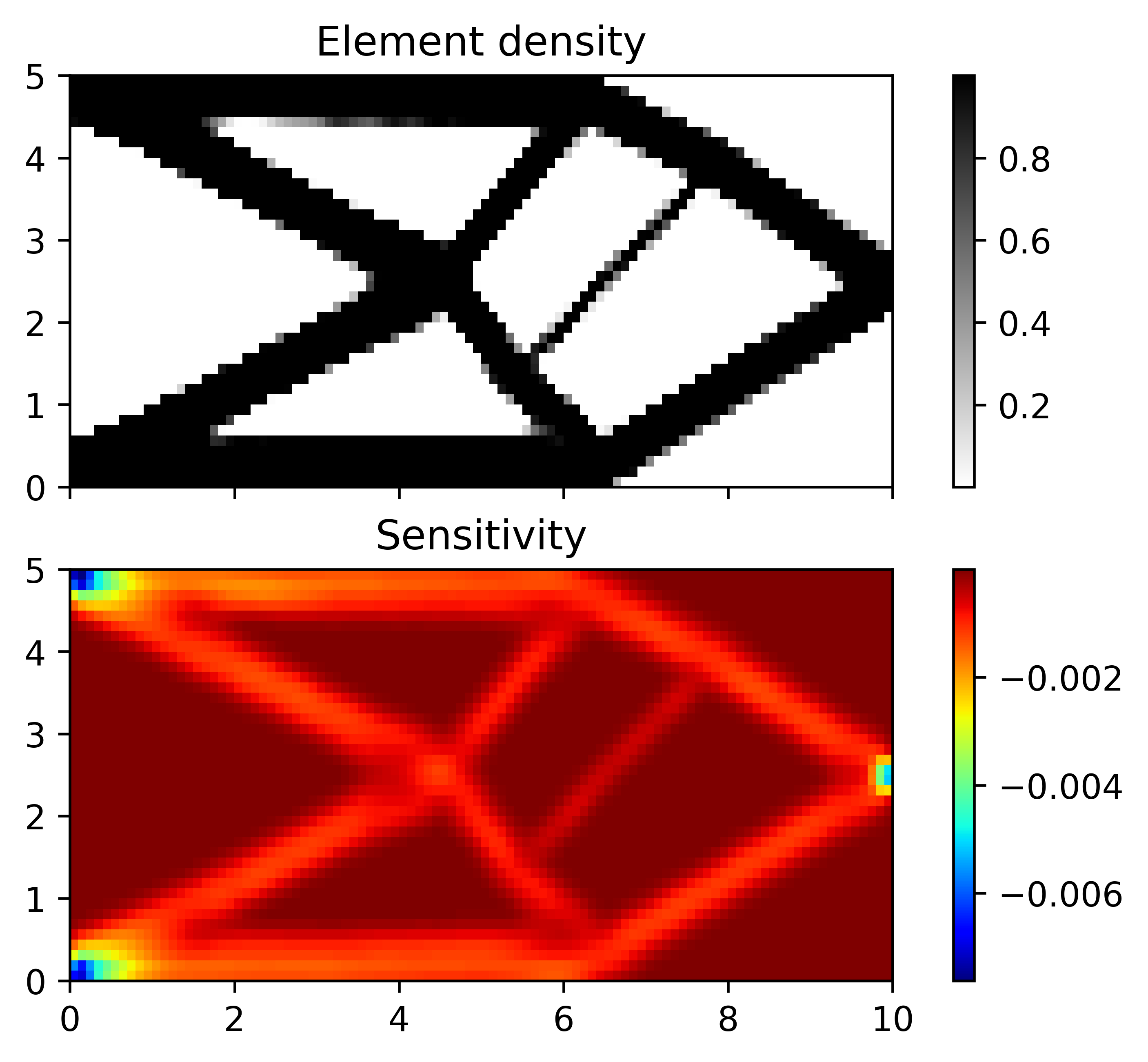}
        \label{fig:d4}}
    \end{minipage}\\
    
    & \\
    
    \begin{minipage}[c]{\x\textwidth}
       \centering 
        \subfloat[FEM, iteration 2]{\includegraphics[trim={0.5cm 5.9cm 2.1cm 0.65cm},clip,width=\textwidth]{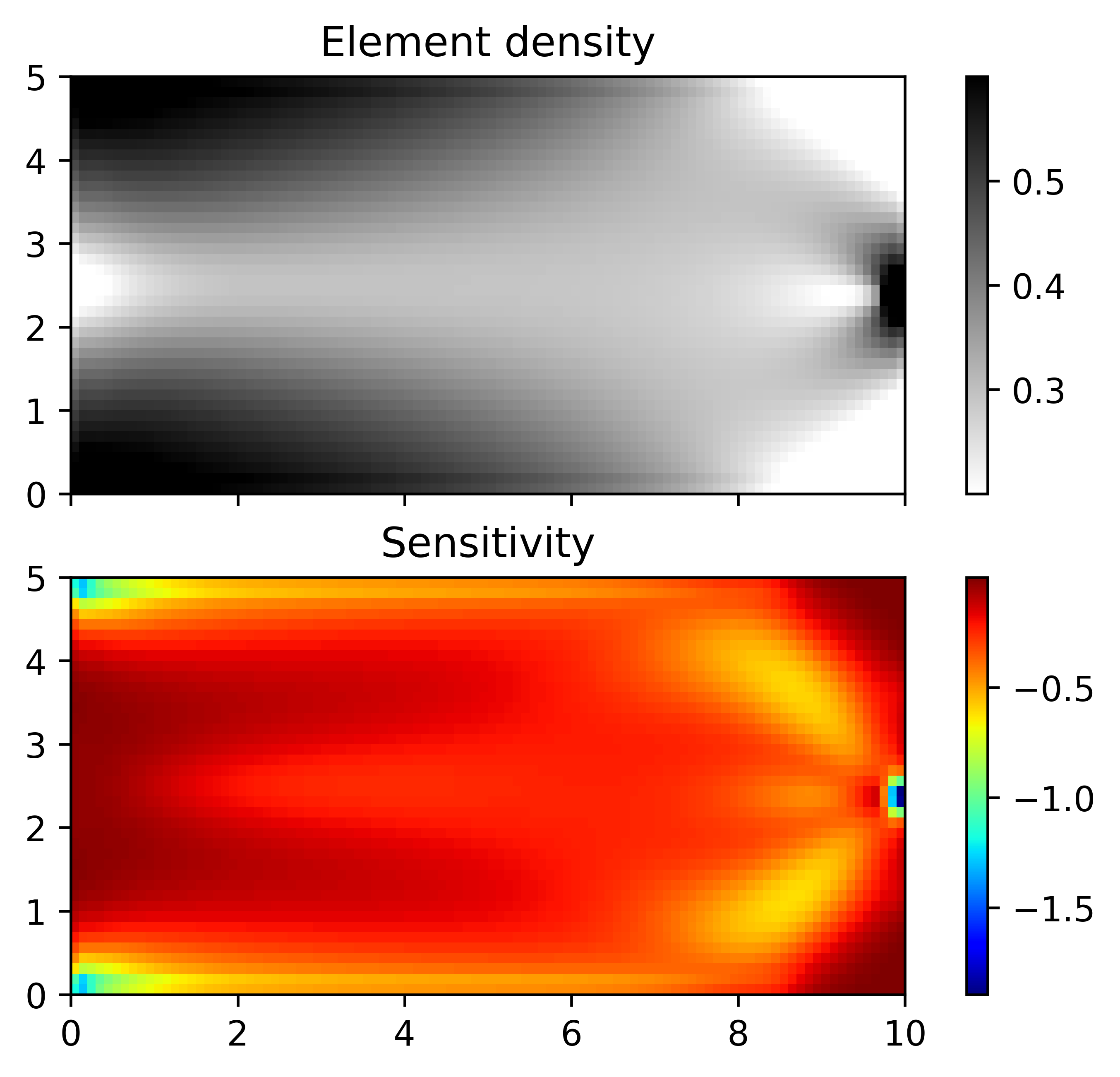}
        \label{fig:d5}}
    \end{minipage}
    &
    \begin{minipage}[c]{\x\textwidth}
       \centering 
        \subfloat[FEM, iteration 15]{\includegraphics[trim={0.5cm 5.9cm 2.1cm 0.65cm},clip,width=\textwidth]{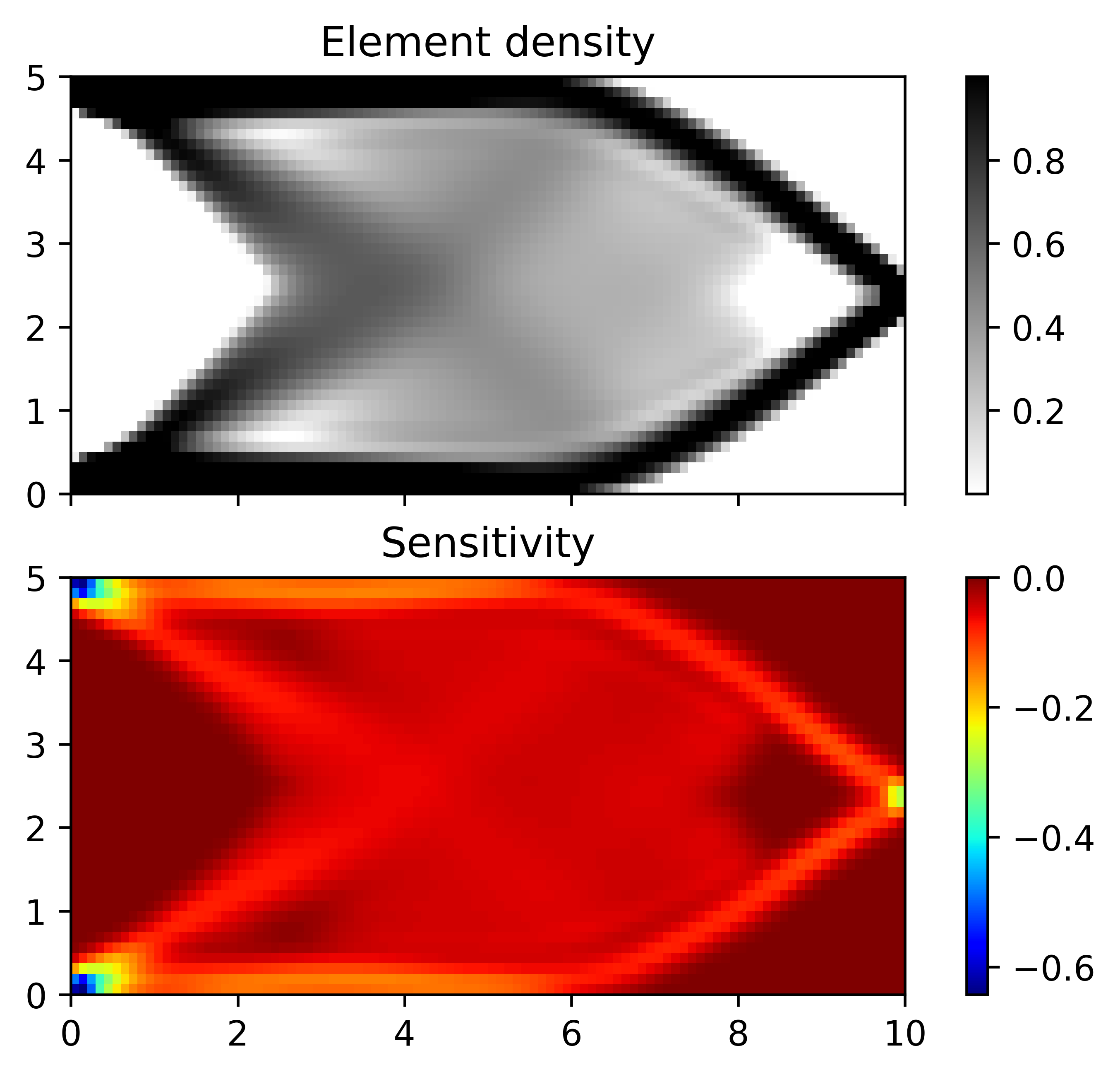}
        \label{fig:d6}}
    \end{minipage}
    &
    \begin{minipage}[c]{\x\textwidth}
       \centering 
        \subfloat[FEM, iteration 30]{\includegraphics[trim={0.5cm 5.9cm 2.1cm 0.65cm},clip,width=\textwidth]{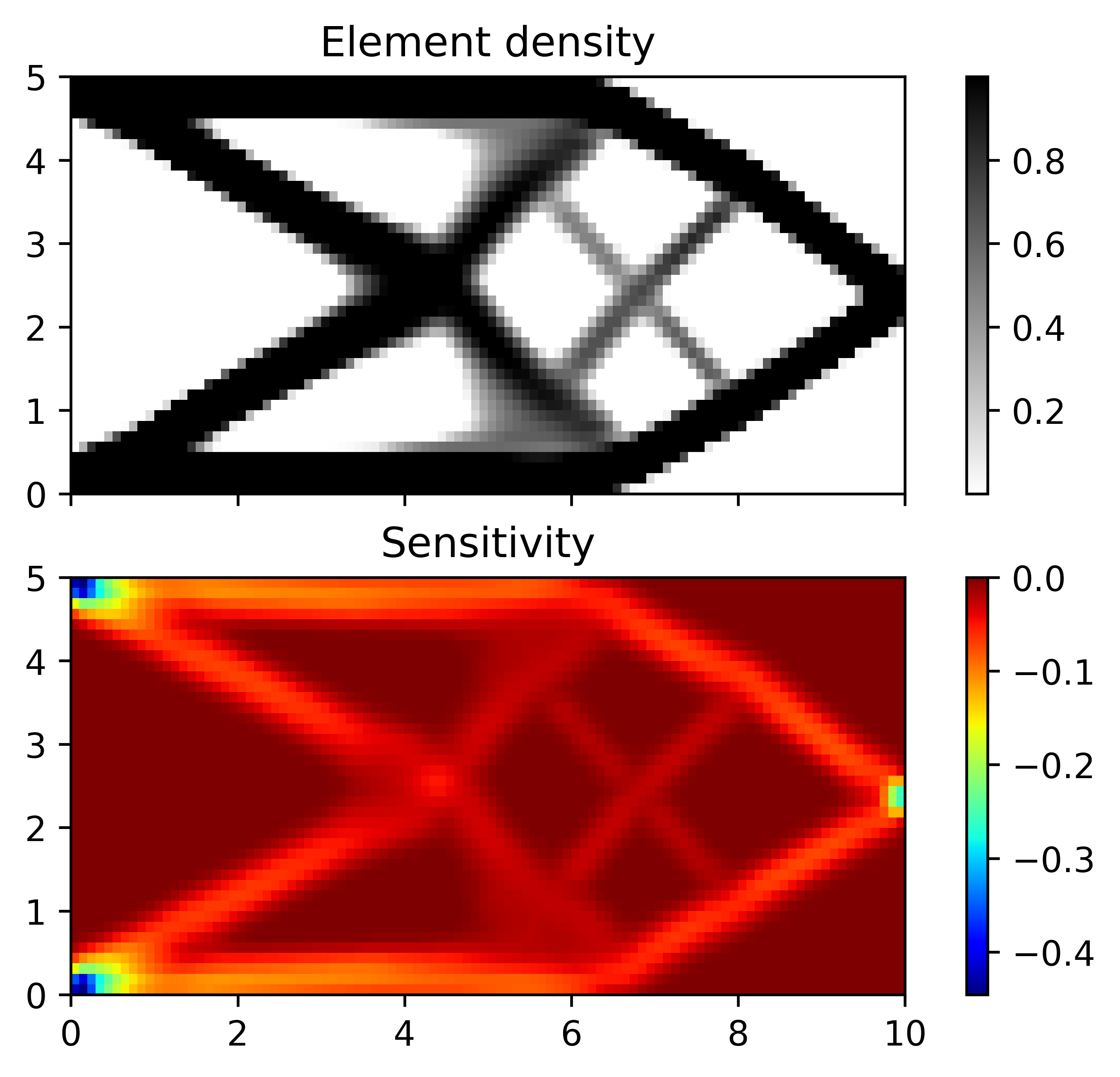}
        \label{fig:d7}}
    \end{minipage}
    &
    \begin{minipage}[c]{\x\textwidth}
       \centering 
        \subfloat[FEM, iteration 80]{\includegraphics[trim={0.5cm 5.9cm 2.1cm 0.65cm},clip,width=\textwidth]{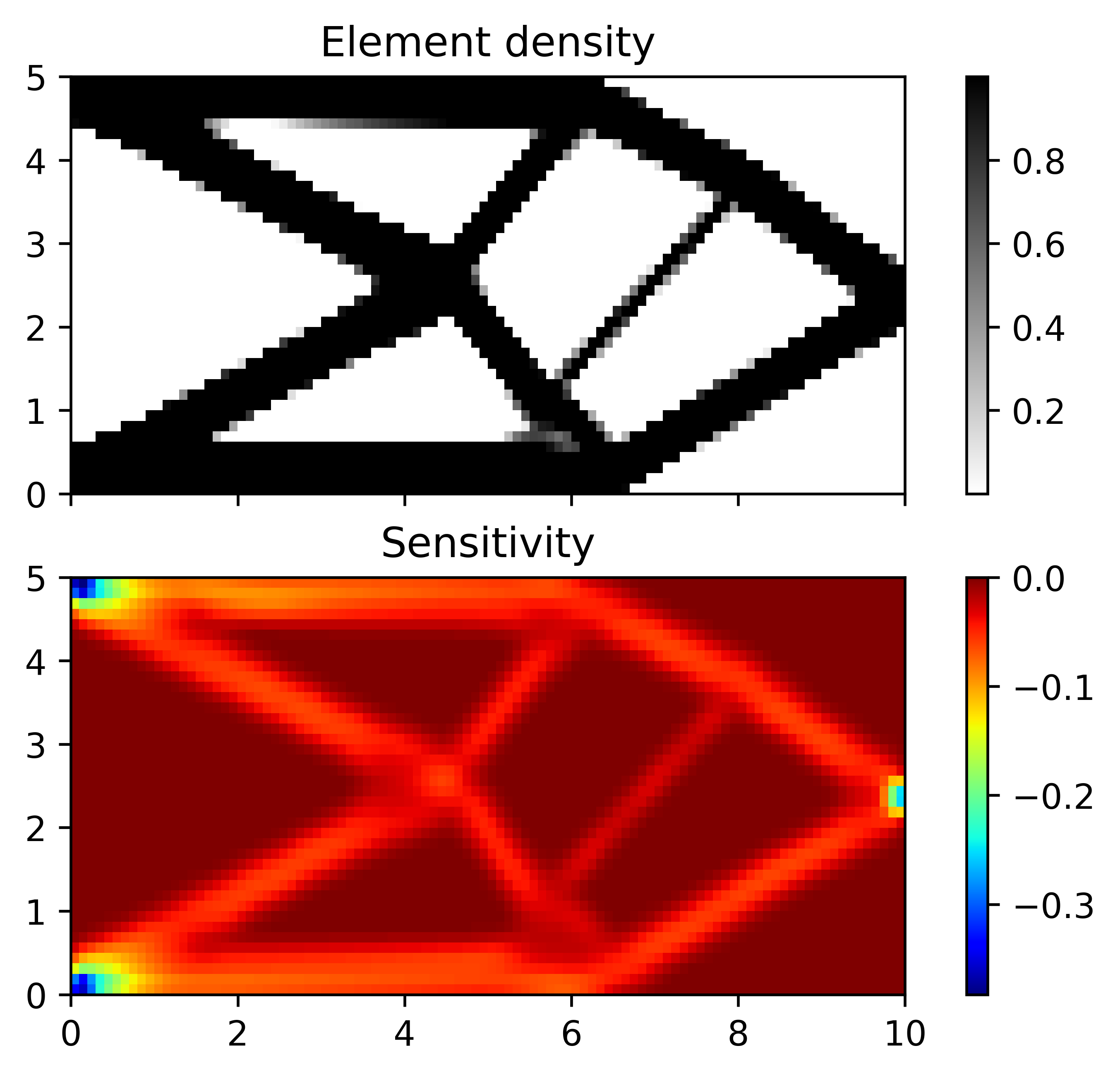}
        \label{fig:d8}}
    \end{minipage}

    \end{tabular}
    \caption{Density plot of designs generated by DEM and FEM for the beam example. The left edge is fixed and a downward load is applied at the center of the right edge.}
    \label{case2_FEM-DEM}
\end{figure}

\begin{figure}[h!] 
    \centering
     \subfloat[]{
         \includegraphics[width=0.33\textwidth]{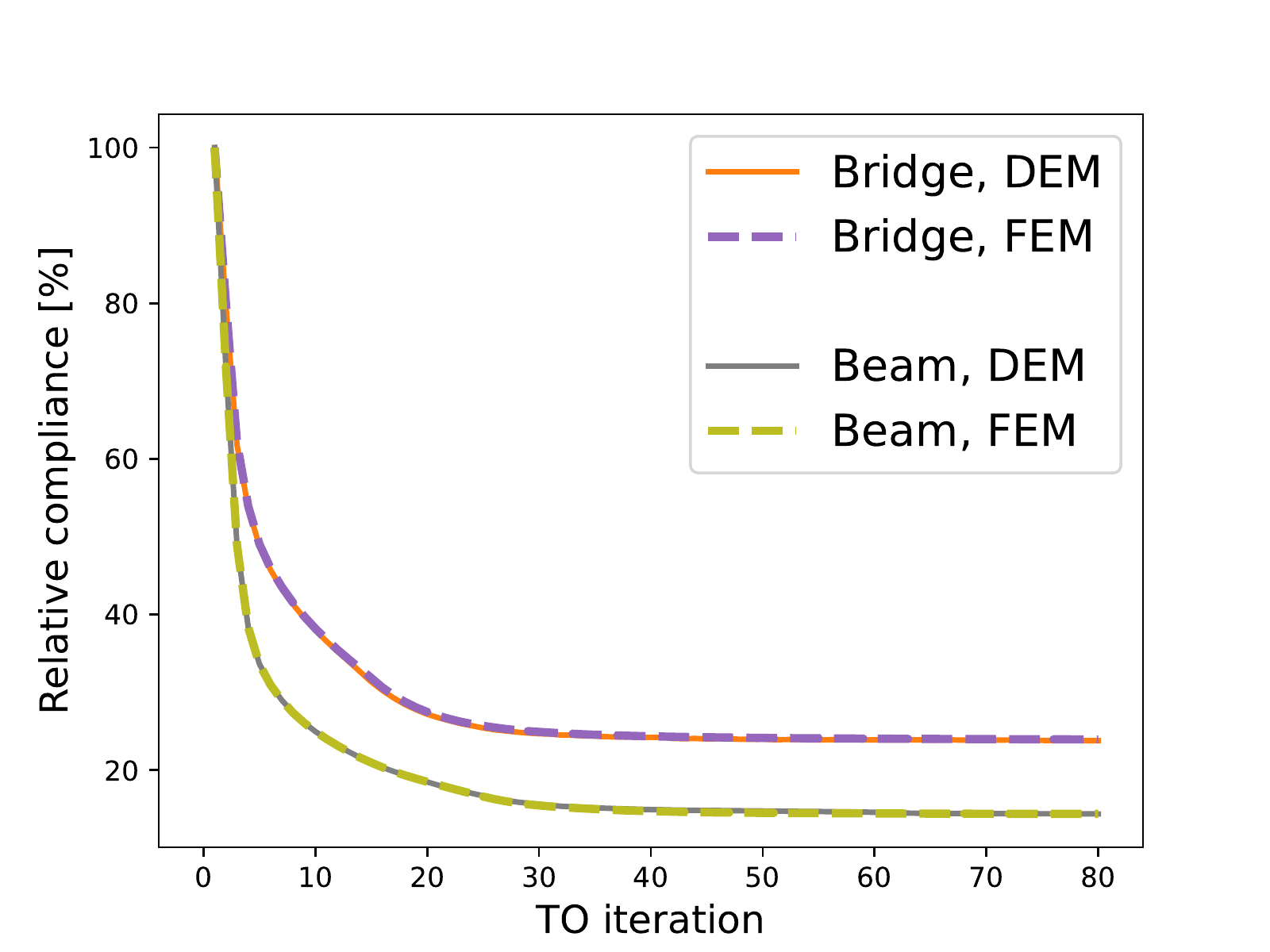}
         \label{fig:compliance_history_2d}
     }
     \subfloat[]{
         \includegraphics[width=0.33\textwidth]{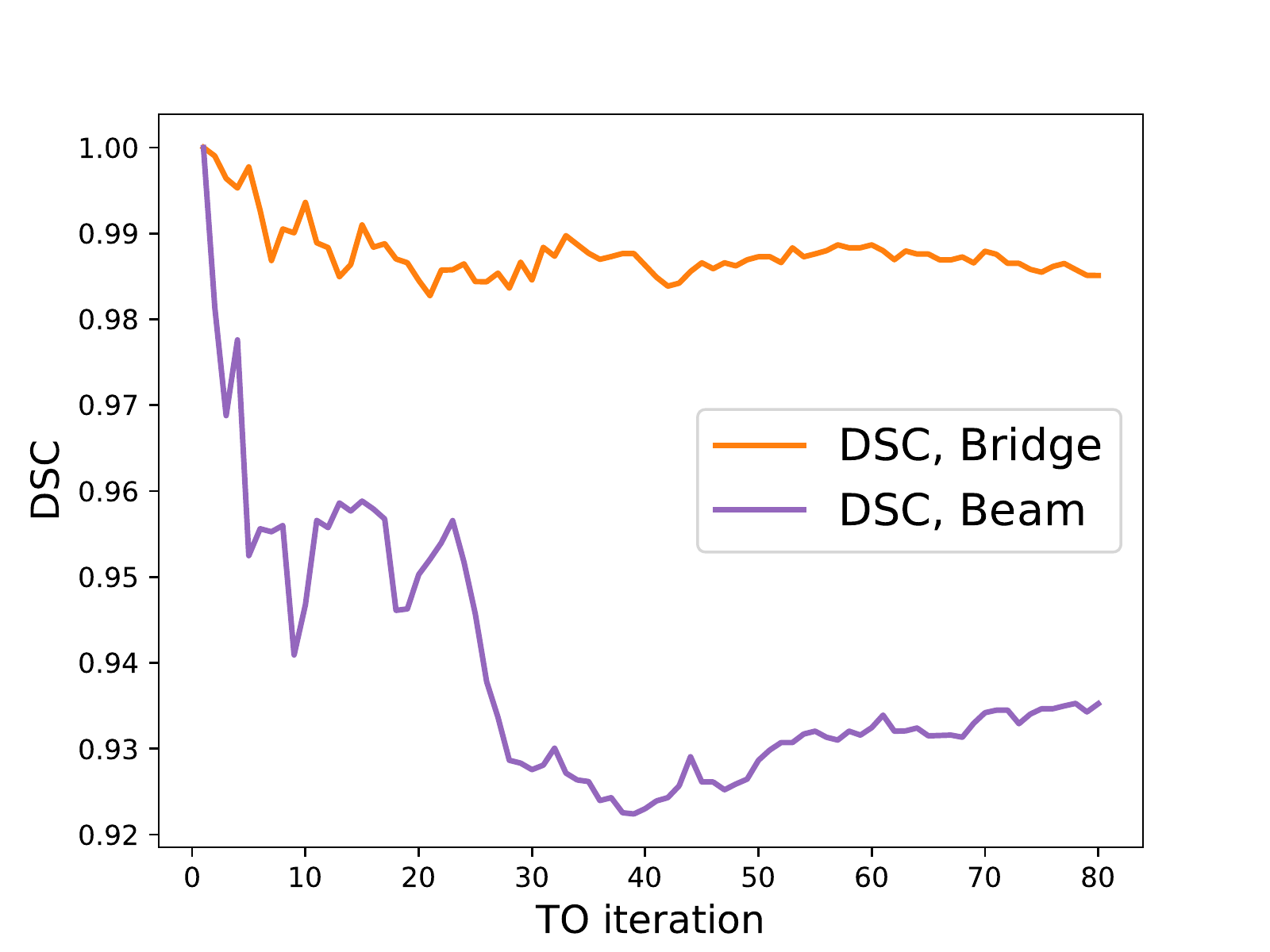}
         \label{fig:DSC}
     }
     \subfloat[]{
         \includegraphics[width=0.33\textwidth]{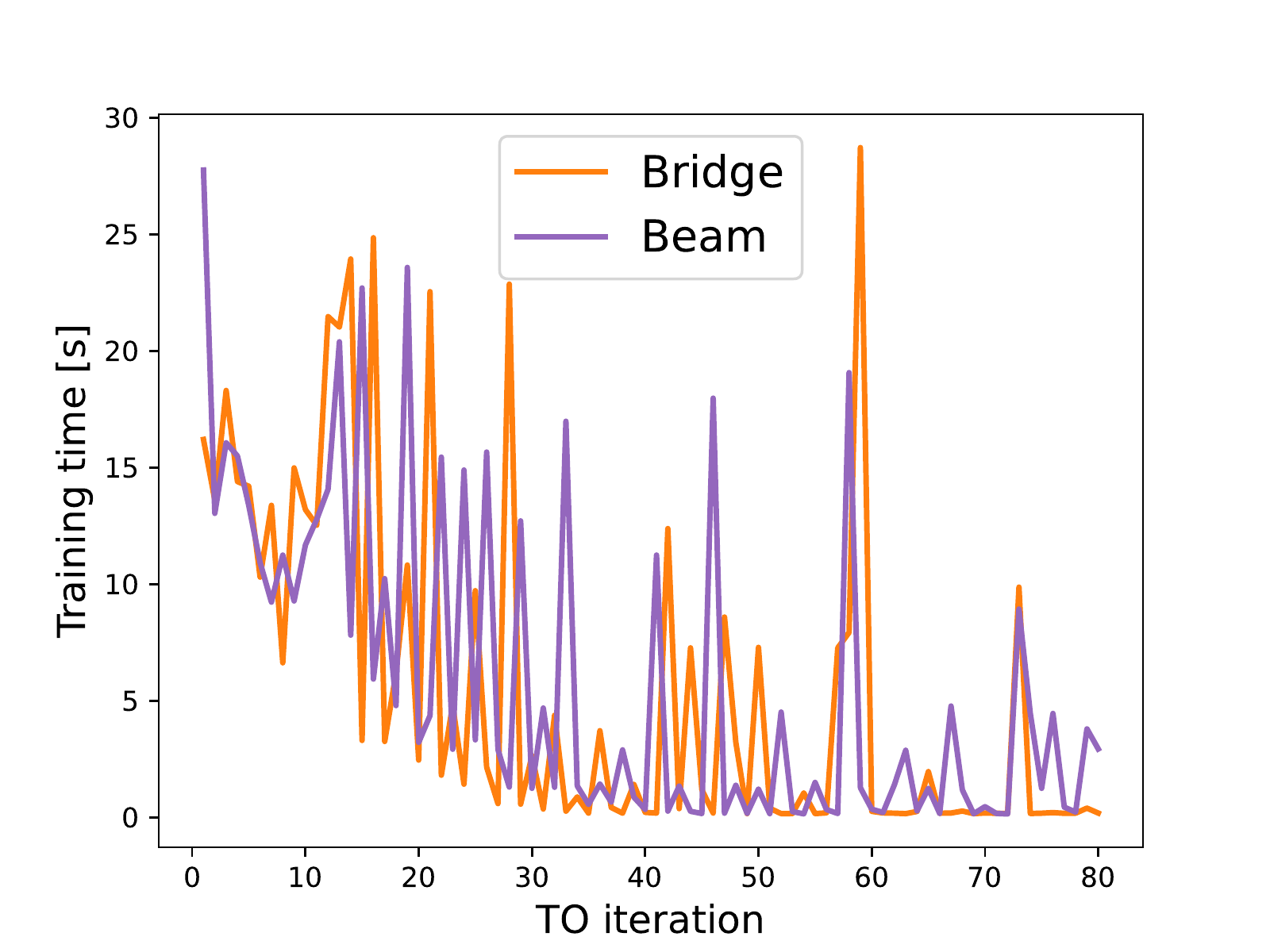}
         \label{fig:train2d}
     }
    \caption{Comparing DEM-based TO with FEM-based TO in 2D: \psubref{fig:compliance_history} Relative compliance evolution for the two cases. \psubref{fig:DSC} Dice similarity coefficient computed on the binarized designs.
    \psubref{fig:train2d} Training time for each iteration.}
\end{figure}

The designs from DEM- and FEM-based TO at different design iterations are compared in Figures \ref{case1_FEM-DEM} and \ref{case2_FEM-DEM}. The relative compliance reduction for both cases as the TO optimizer progressed is presented in \fref{fig:compliance_history_2d}. To give a quantitative comparison of the similarity between DEM- and FEM-based designs, the dice similarity coefficient (DSC) was computed for both cases and is presented in \fref{fig:DSC}. The DSC is defined as:
\begin{equation}
    {\rm{DSC}} = \frac{ 2 | \bm{I}_{FEM} * \bm{I}_{DEM} |}{ |\bm{I}_{FEM}| + |\bm{I}_{DEM}|},
    \label{eq:DSC_eq}
\end{equation} 
where $\bm{I}$ represents the binarized density using a (dynamic) threshold of 0.5 times the current maximum density. Finally, the computational cost for running the two cases using DEM and FEM\footnote{The highly optimized Python TO code found in \cite{pyTO} was used for time comparison, which was based on the MATLAB code by Andreassen et al. \cite{andreassen2011efficient}} are summarized in \fref{fig:train2d} and are compared in \tref{ex1_time}.

\begin{table}[h!]
    \caption{Computational time for the two cases in Example 1}
    \small
    \centering
    \begin{tabular}{cccc}
      & \vline & FEM-based  & DEM-based \\
    \hline
    Bridge & \vline & 4.1s  & 460.7s \\
    \hline
    Beam & \vline & 12.7s  & 485.7s \\
    \hline
    \end{tabular}
    \label{ex1_time}
\end{table}

From \fref{case1_FEM-DEM} and \fref{case2_FEM-DEM}, we clearly see that the DEM- and FEM-based TO frameworks produced almost identical simulation design evolution history and final designs. The similarity in designs is further demonstrated in \fref{fig:DSC}, where we see that the DSC is greater than 90\% for the final designs in both cases. For the bridge case, we noticed that the DSC remains greater than 98\% for all design iterations, showing close resemblance between the DEM- and FEM-based approaches. The two frameworks also produce designs of similar performance, as is evident in \fref{fig:compliance_history_2d}, where we see that the compliance reduction history is almost identical between the two methods due to the use of identical MMA optimizer. Nonetheless, \tref{ex1_time} shows that FEM still holds a massive advantage over DEM in terms of computational time, which agrees with the trends observed in the similar DEM-based TO work like \cite{zehnder2021ntopo}. However, a very tight relative tolerance of $5\times10^{-6}$ was used in training, which could be loosened to gain computational efficiency. From \fref{fig:train2d}, we highlight two important features of DEM-based TO. For both cases, we noticed that the training time shows a decreasing trend as design progressed. This is due to the fact that the weights and biases from the previous TO iteration are used as the initial condition for the next iteration, a form of transfer learning. This leads to gradual training time reduction. This is in sharp contrast of FEM-based TO, where the computational time for each iteration is constant, and can be very beneficial when many TO iterations are needed. Also, we note that the DEM training time went up by merely 5.4\% from case 1 to case 2, where the FEM-based TO time went up over 200\%. The insensitivity of training time to domain size is due to the fact that the number of parameters of DEM-based TO is controlled by the DNN architecture, \emph{not} by the number of nodes used to discretize the domain, which is beneficial for problems that need to be resolved by a fine mesh.

\subsection{Compliance minimization in 3D}
\label{3d_comp}
The second example concerns a 3D rectangular bridge of size 12-by-2-by-2 m$^3$. It was fixed at both end surfaces and was subjected to a downward load at the center of the top surface in a square region of size 0.5-by-0.5 m$^2$. 75625 nodes were placed into the domain, forming a 121-by-25-by-25 grid. $\epsilon_{tol}$ was set to $5\times10^{-5}$. As comparison, Abaqus/Standard \cite{Abaqus2021} was used to perform TO using the built-in TO functionalities and optimizer.

\begin{figure}[b!]
    \newcommand\x{0.45}
    \centering
    \begin{tabular}{ c  c   }
    \begin{minipage}[c]{\x\textwidth}
       \centering                                         
        \subfloat[DEM, iteration 80]{\includegraphics[trim={16cm 9cm 20cm 3cm},clip,width=\textwidth]{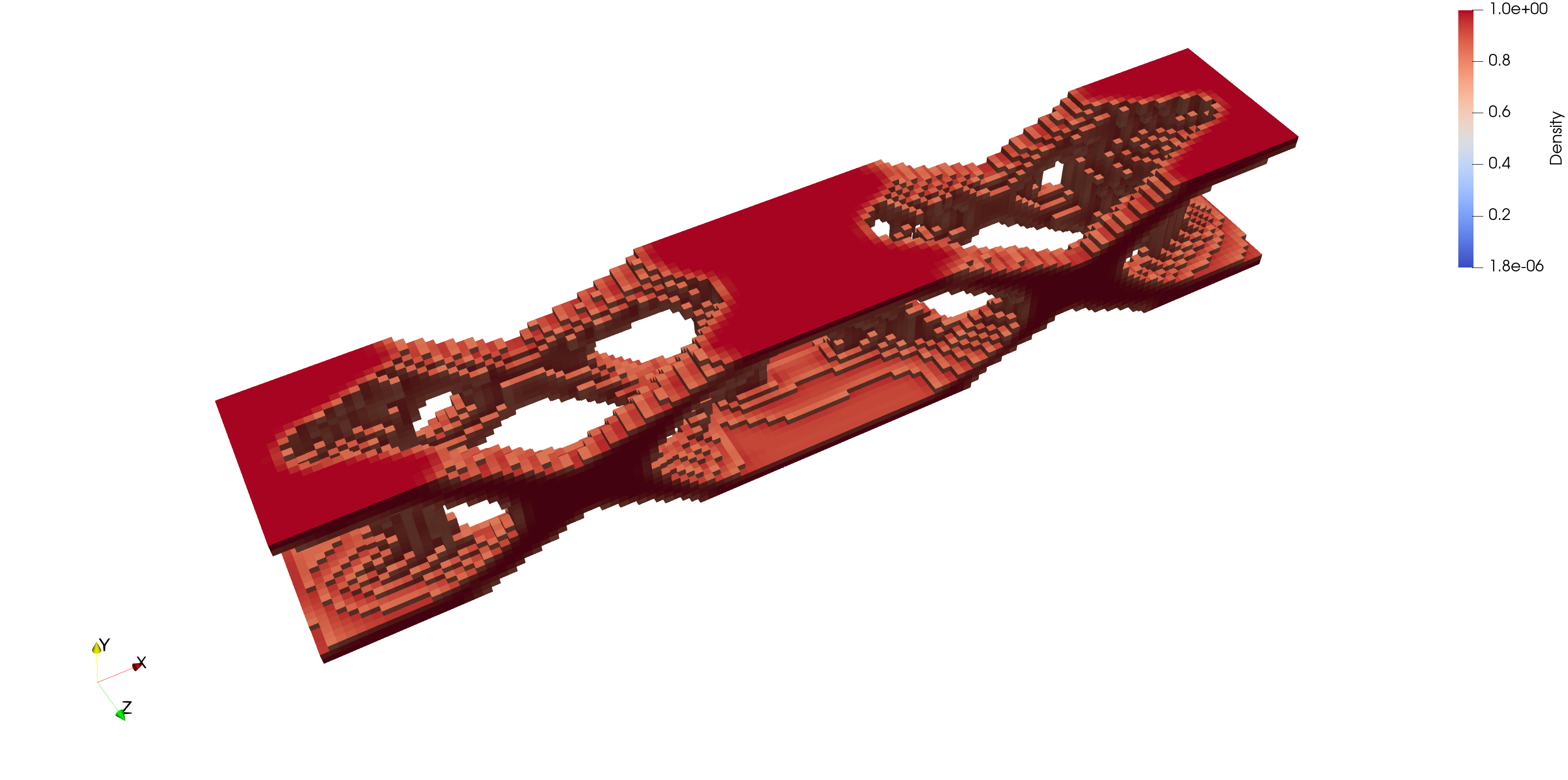}
        \label{fig:dem_final}}
    \end{minipage}
    &
    \begin{minipage}[c]{\x\textwidth}
       \centering 
        \subfloat[FEM, iteration 80]{\includegraphics[trim={16cm 9cm 20cm 3cm},clip,width=\textwidth]{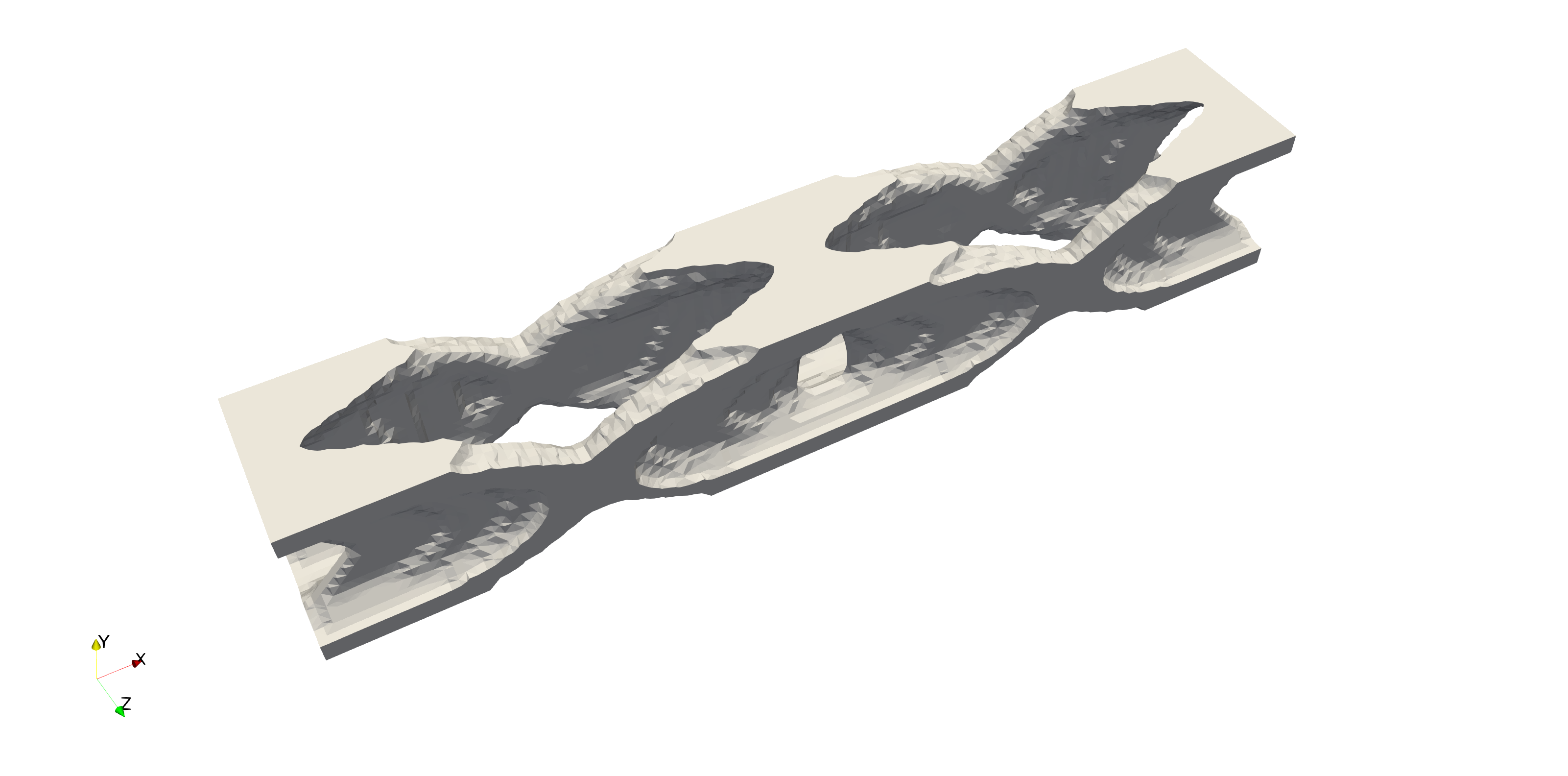}
        \label{fig:fem_final}}
    \end{minipage}\\
    
    & \\
    
    \begin{minipage}[c]{\x\textwidth}
       \centering 
        \subfloat[Overlapping two designs]{\includegraphics[trim={16cm 9cm 20cm 3cm},clip,width=\textwidth]{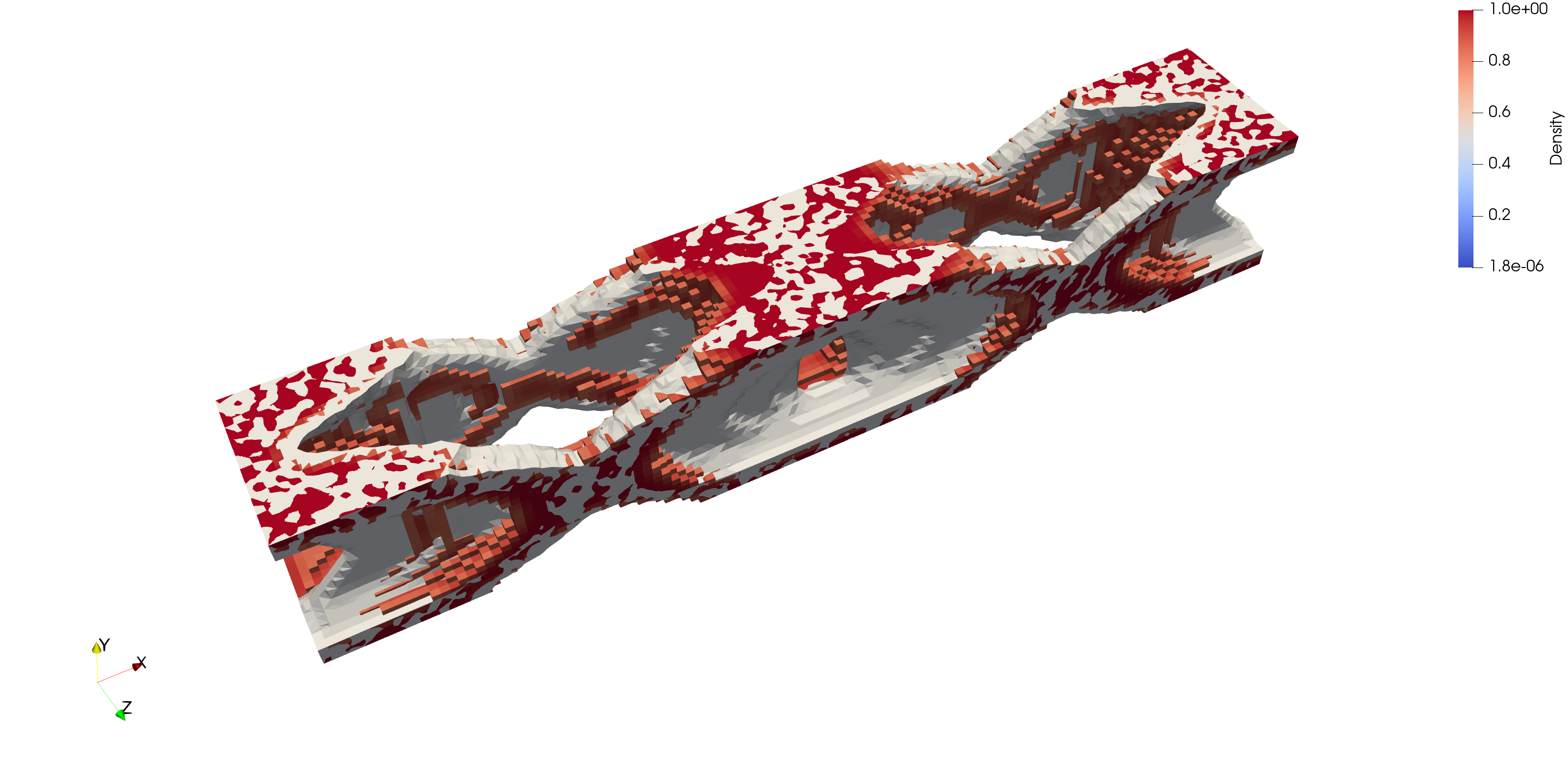}
        \label{fig:over1}}
    \end{minipage}
    &
    \begin{minipage}[c]{\x\textwidth}
       \centering 
        \subfloat[Section view]{\includegraphics[trim={16cm 9cm 20cm 3cm},clip,width=\textwidth]{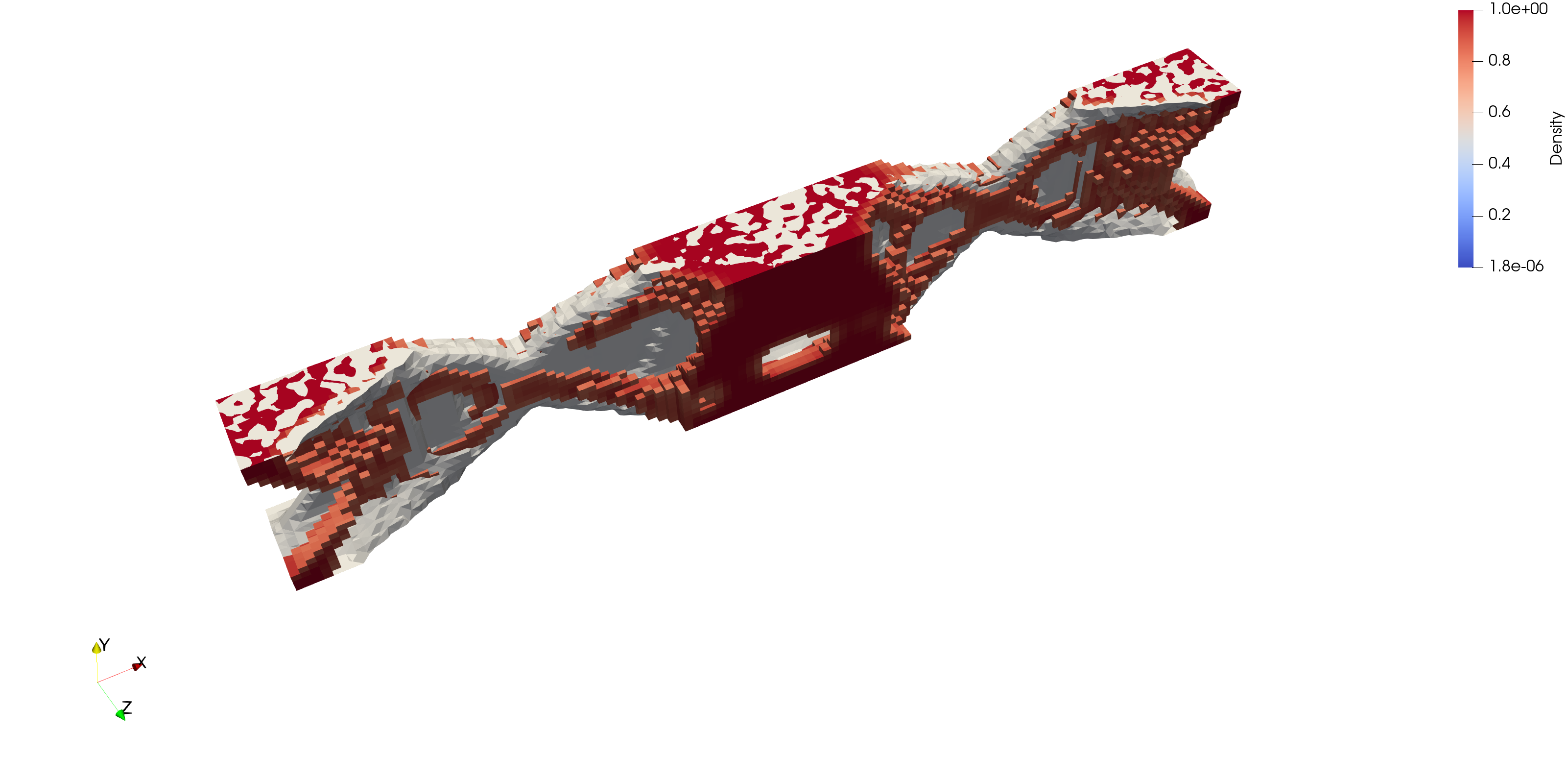}
        \label{fig:over2}}
    \end{minipage}

    \end{tabular}
    \caption{Comparison of final designs after 80 design iterations. The FEM-based design is shown in the form of a stl surface, which the DEM design is shown as voxels.}
    \label{Ex3D_FEM-DEM}
\end{figure}

\begin{figure}[h!] 
    \centering
     \subfloat[]{
         \includegraphics[width=0.5\textwidth]{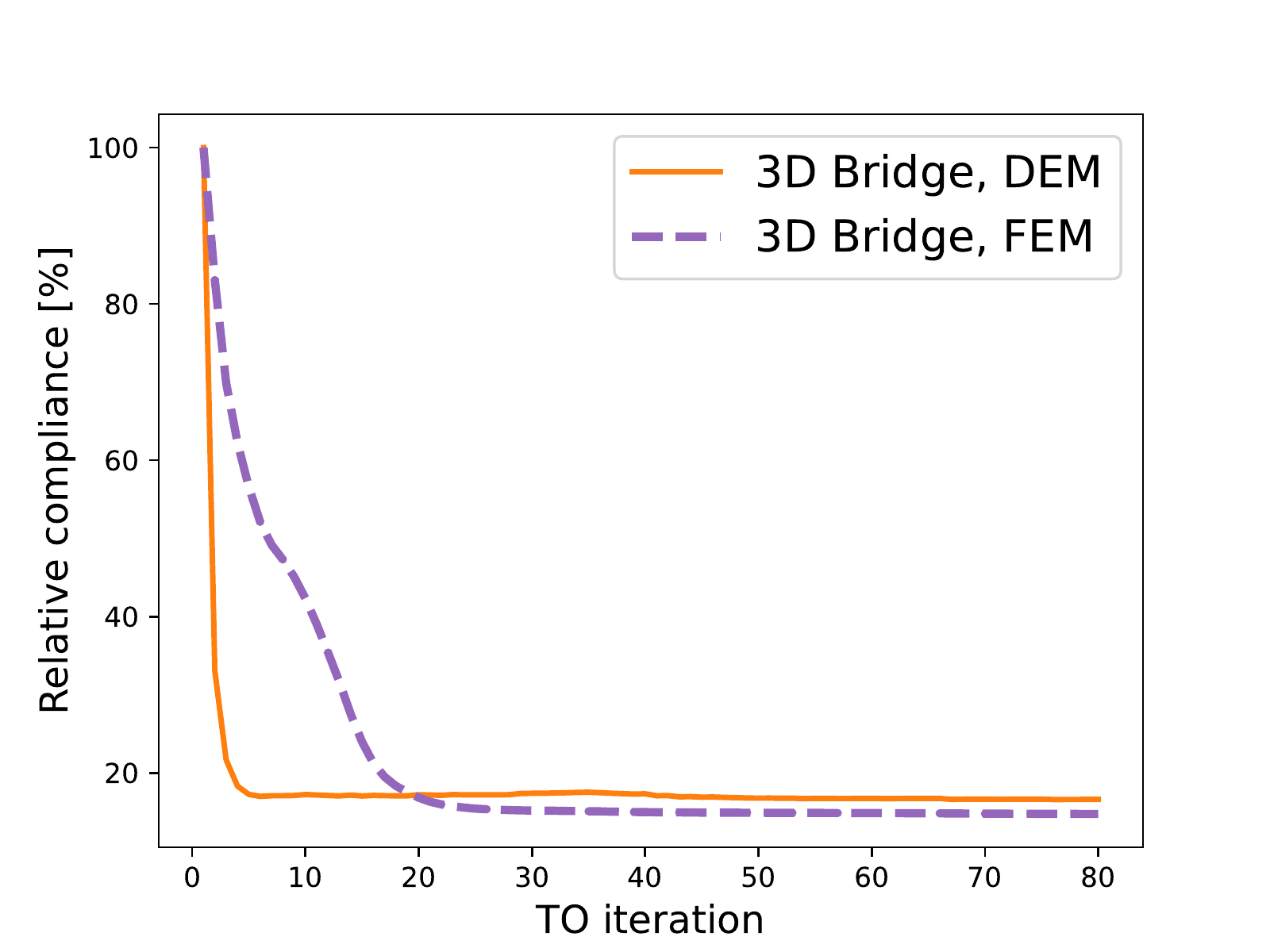}
         \label{fig:compliance_history}
     }
     \subfloat[]{
         \includegraphics[width=0.5\textwidth]{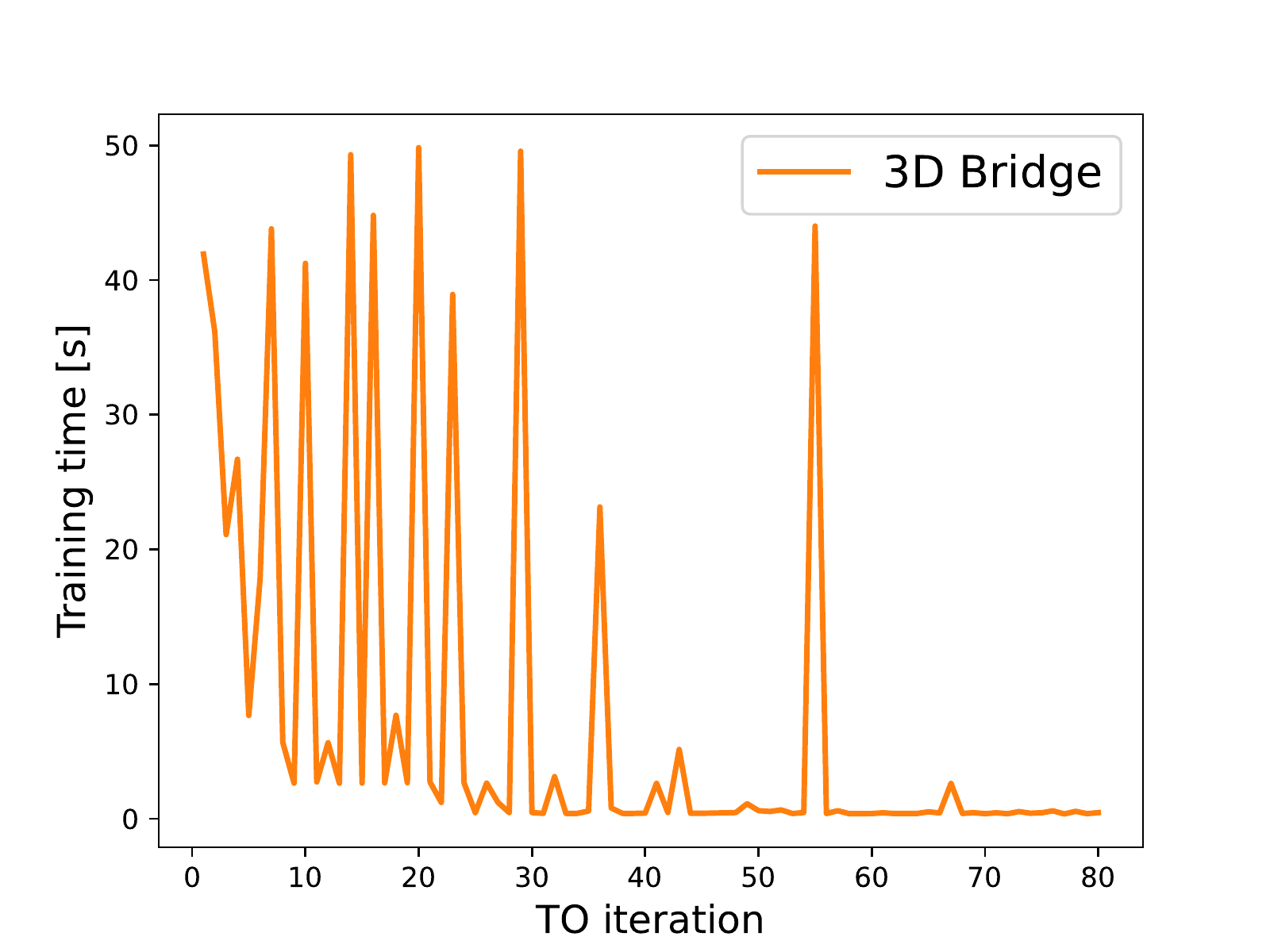}
         \label{fig:train3d}
     }
    \caption{DEM-based TO in 3D: \psubref{fig:compliance_history} Relative compliance evolution for the 3D bridge design.
    \psubref{fig:train3d} Training time for each iteration.}
    \label{Ex3D_compliance}
\end{figure}

The final designs from DEM and FEM are presented in \fref{fig:dem_final} and \fref{fig:fem_final}, respectively. The STL file for the final FEM design was extracted using a density threshold of 0.8, and the same threshold was used to obtain the voxelated DEM final design. The two designs are overlaid and presented in \fref{fig:over1} and \fref{fig:over2}. The relative compliance reduction and training time was shown in \fref{Ex3D_compliance}. The total TO time for DEM was 1050.4s.

From the overlaid plot and the section view in \fref{Ex3D_FEM-DEM}, we see that the DEM-based design largely resembles the FEM-based design, especially near both ends of the bridge. Some design differences can be seen in the center, especially from the section view. From \fref{fig:compliance_history}, we see that DEM was able to quickly reduce the compliance in less than 10 iterations, but has a final compliance that is larger than that of the FEM design. The evolution histories are different, which roots from the fact that different optimizers were used to optimize the material distribution. From \fref{fig:train3d}, we again see that training time decreased as design progressed, a trend that is consistent with \sref{2d_comp}. Finally, we highlight that the same hyperparameters in \sref{2d_comp} were used directly in this 3D example without any additional hyperparameter optimization, which shows the robustness of the DEM model with respect to hyperparameter selection.

\subsection{Shear modulus maximization in 2D}
\label{shear_2D}
\begin{figure}[h!]
    \newcommand\x{0.3}
    \centering
    \begin{tabular}{ c  c  c  }
    \begin{minipage}[c]{\x\textwidth}
       \centering                                         
        \subfloat[Initial, r=H/4]{\includegraphics[trim={0.9cm 5.9cm 2.7cm 1cm},clip,width=\textwidth]{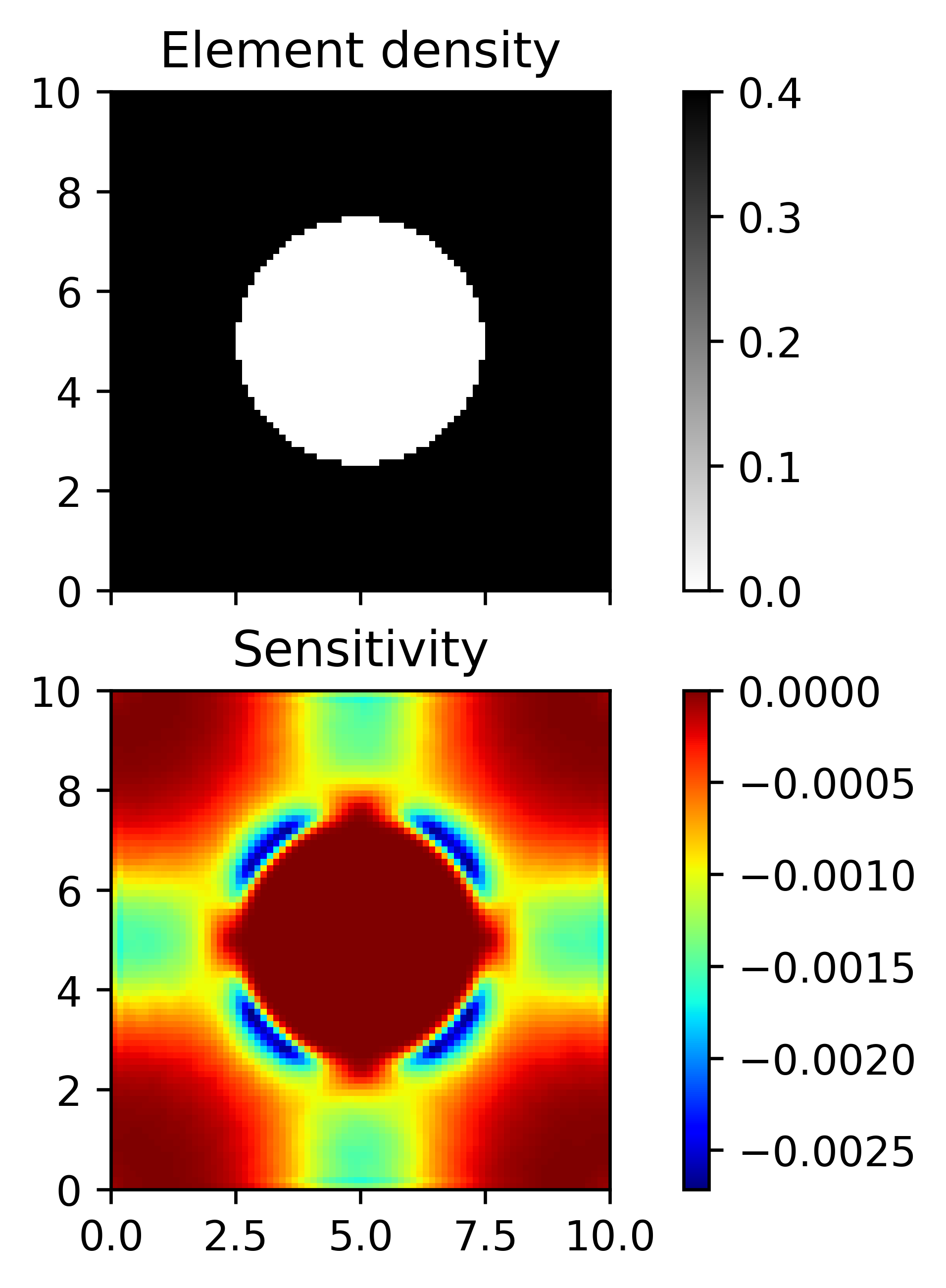}
        \label{fig:d1_}}
    \end{minipage}
    &
    \begin{minipage}[c]{0.29\textwidth}
       \centering 
        \subfloat[Initial, r=H/10]{\includegraphics[trim={0.9cm 5.9cm 2.7cm 1cm},clip,width=\textwidth]{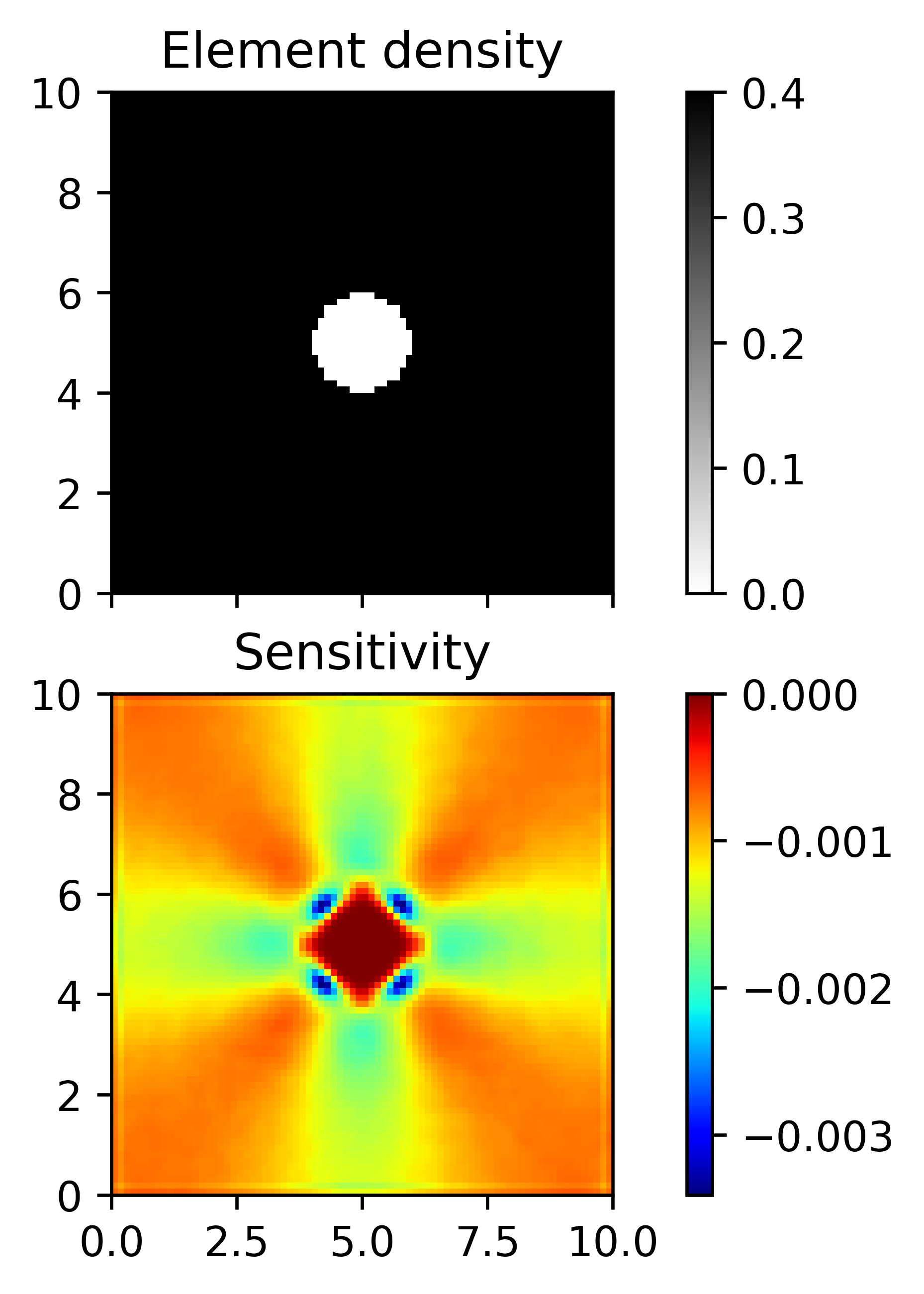}
        \label{fig:d2_}}
    \end{minipage}
    &
    \begin{minipage}[c]{\x\textwidth}
       \centering 
        \subfloat[Initial, r=H/20]{\includegraphics[trim={0.9cm 5.9cm 2.7cm 1cm},clip,width=\textwidth]{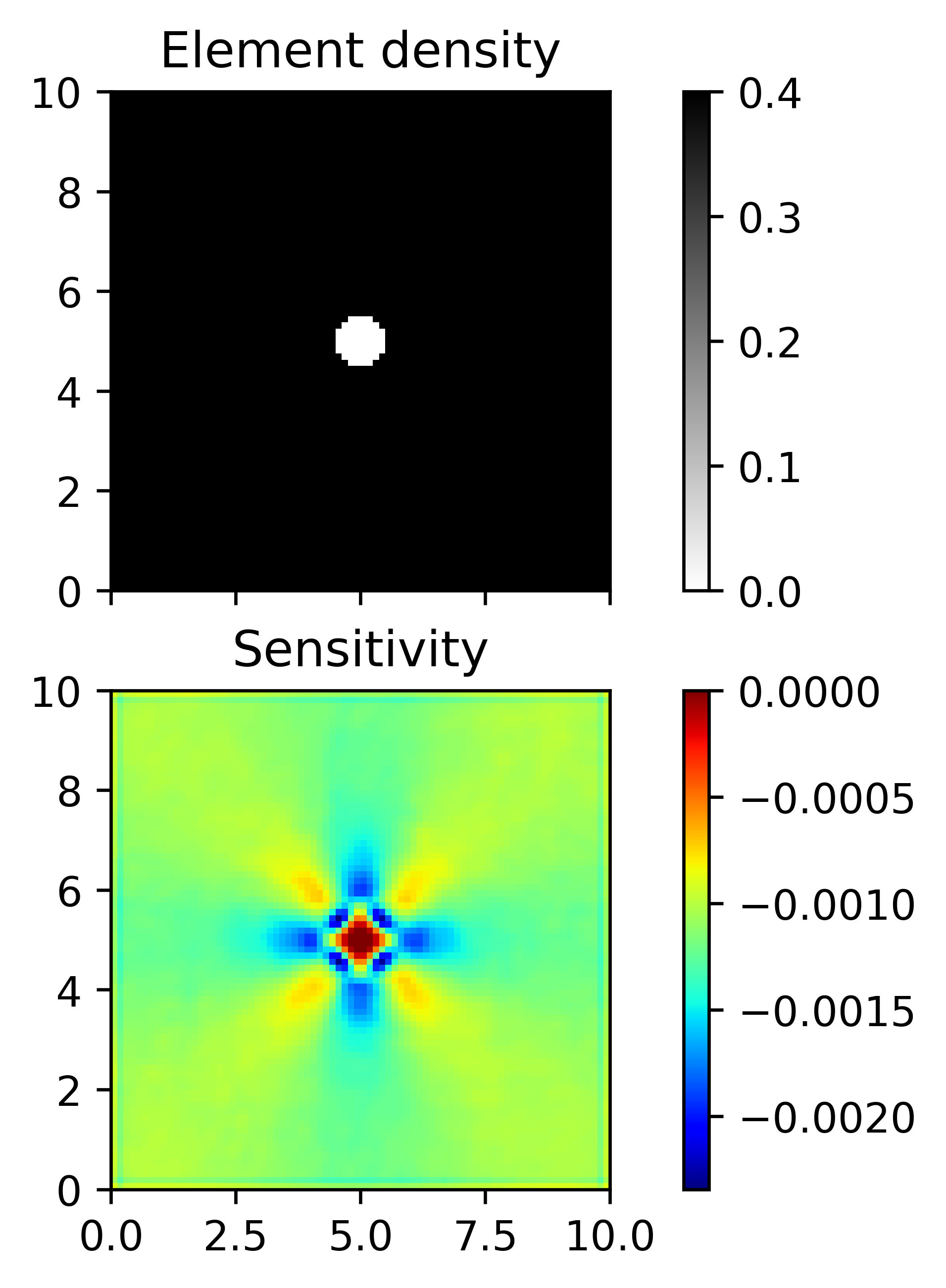}
        \label{fig:d3_}}
    \end{minipage}\\
    
    & \\
    
    \begin{minipage}[c]{0.29\textwidth}
       \centering 
        \subfloat[DEM, iteration 80, r=H/4]{\includegraphics[trim={0.8cm 5.9cm 2.5cm 0.7cm},clip,width=\textwidth]{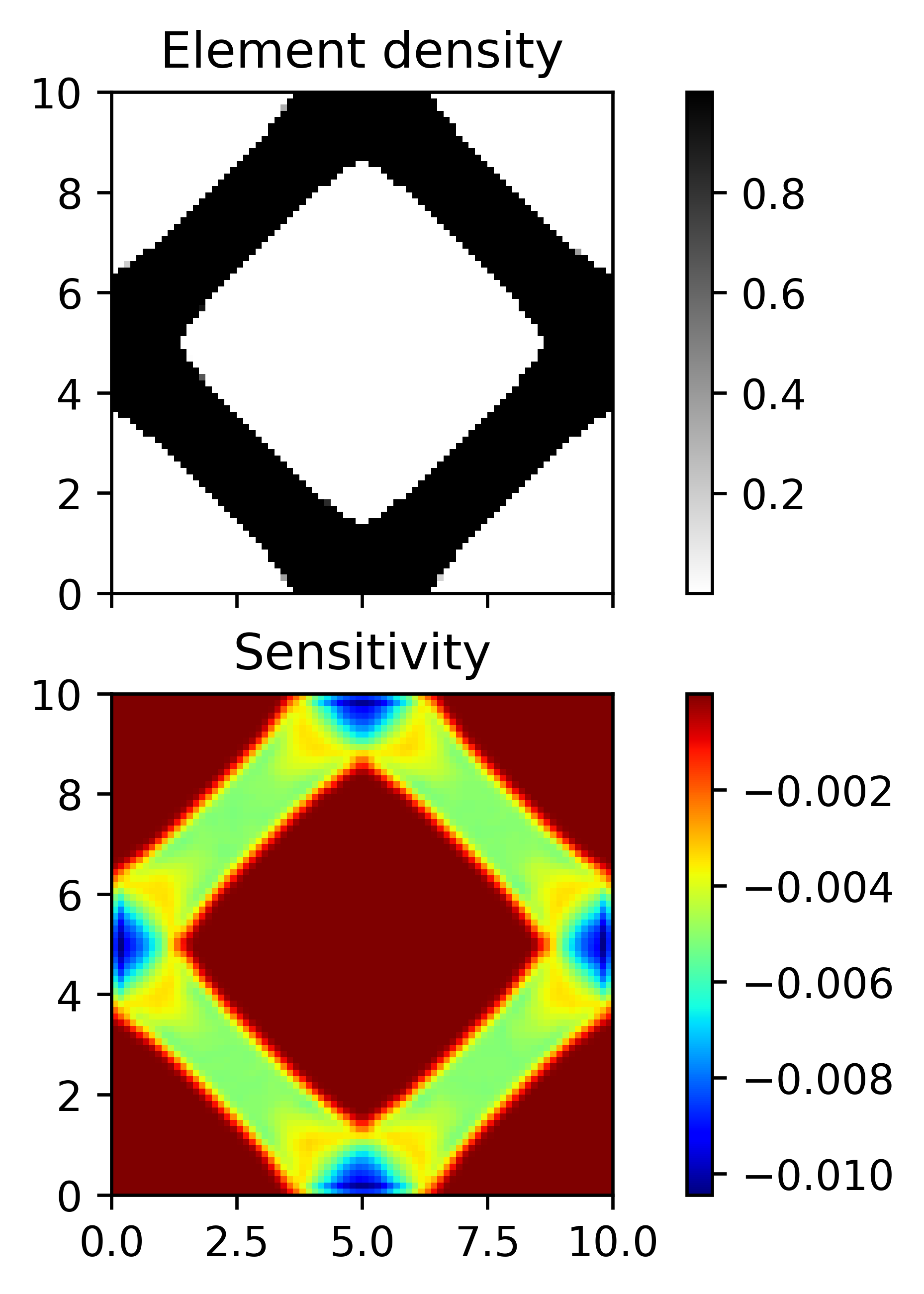}
        \label{fig:d4_}}
    \end{minipage}
    &
    \begin{minipage}[c]{0.29\textwidth}
       \centering 
        \subfloat[DEM, iteration 80, r=H/10]{\includegraphics[trim={0.8cm 5.9cm 2.5cm 0.7cm},clip,width=\textwidth]{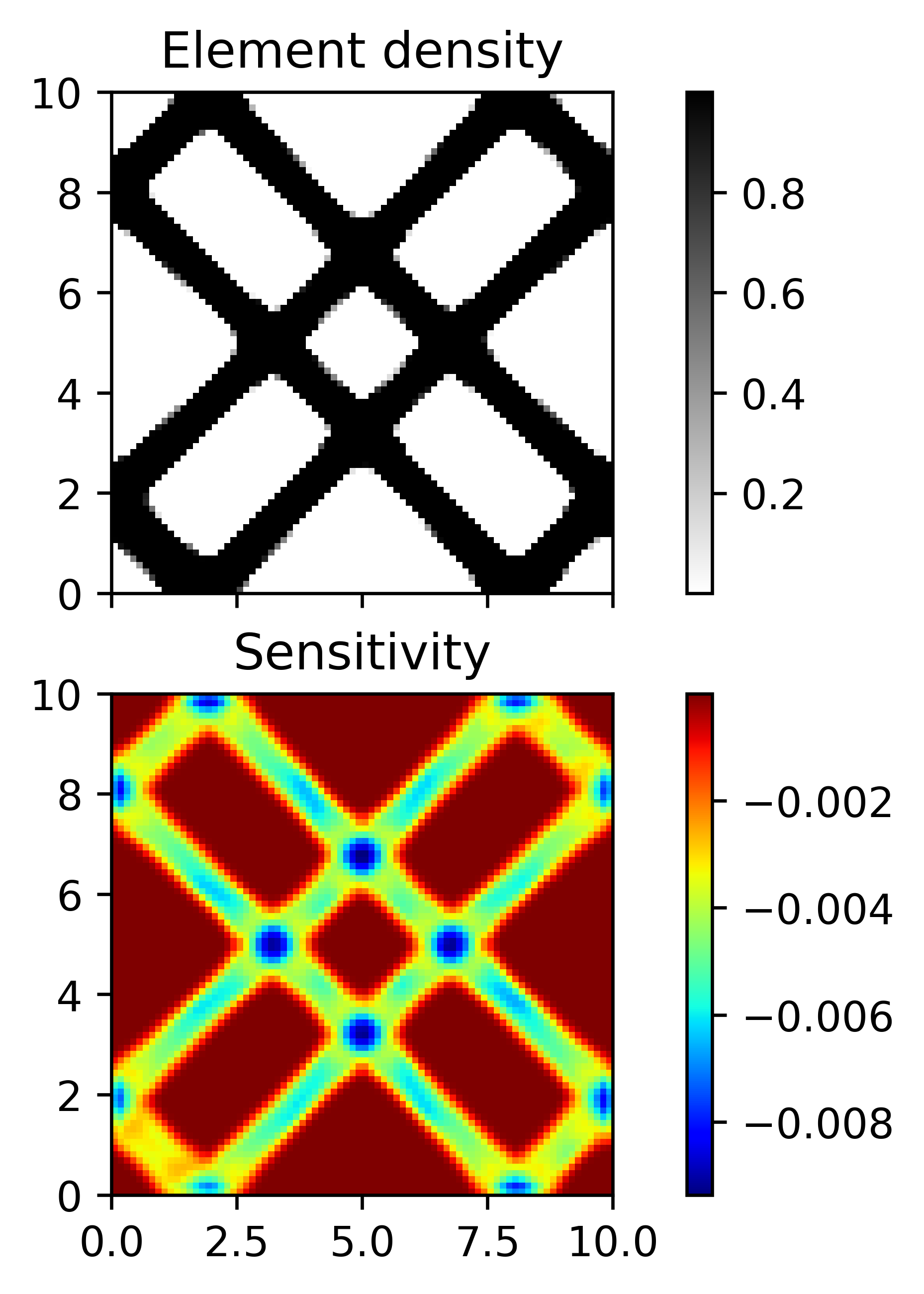}
        \label{fig:d5_}}
    \end{minipage}
    &
    \begin{minipage}[c]{0.29\textwidth}
       \centering 
        \subfloat[DEM, iteration 80, r=H/20]{\includegraphics[trim={0.8cm 5.9cm 2.5cm 0.7cm},clip,width=\textwidth]{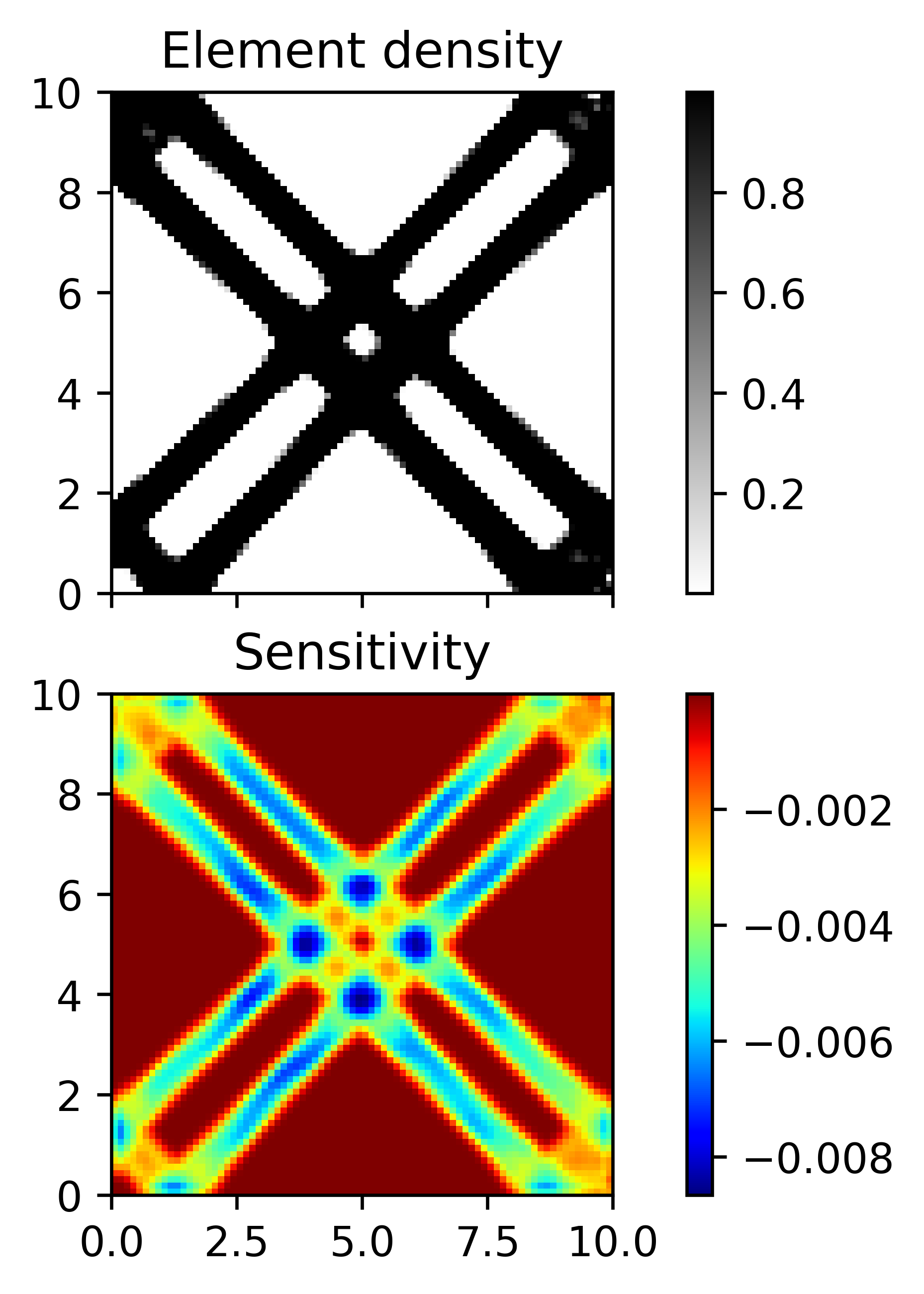}
        \label{fig:d6_}}
    \end{minipage}\\

    & \\
    
    \begin{minipage}[c]{\x\textwidth}
       \centering 
        \subfloat[FEM, iteration 80, r=H/4]{\includegraphics[trim={1.6cm 0.5cm 0.5cm 1.6cm},clip,width=\textwidth]{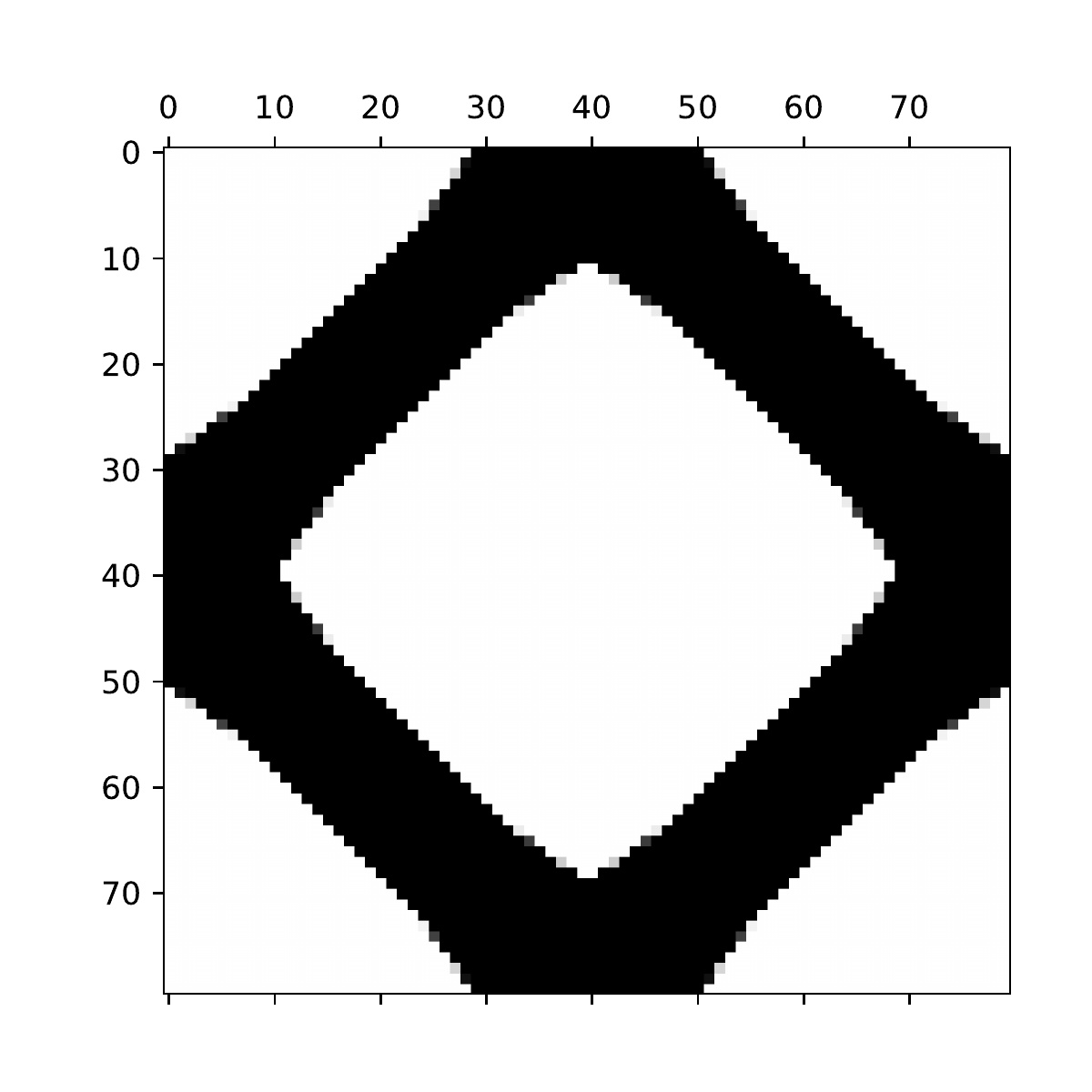}
        \label{fig:d7_}}
    \end{minipage}
    &
    \begin{minipage}[c]{0.285\textwidth}
       \centering 
        \subfloat[FEM, iteration 80, r=H/10]{\includegraphics[trim={1.6cm 0.5cm 0.5cm 1.6cm},clip,width=\textwidth]{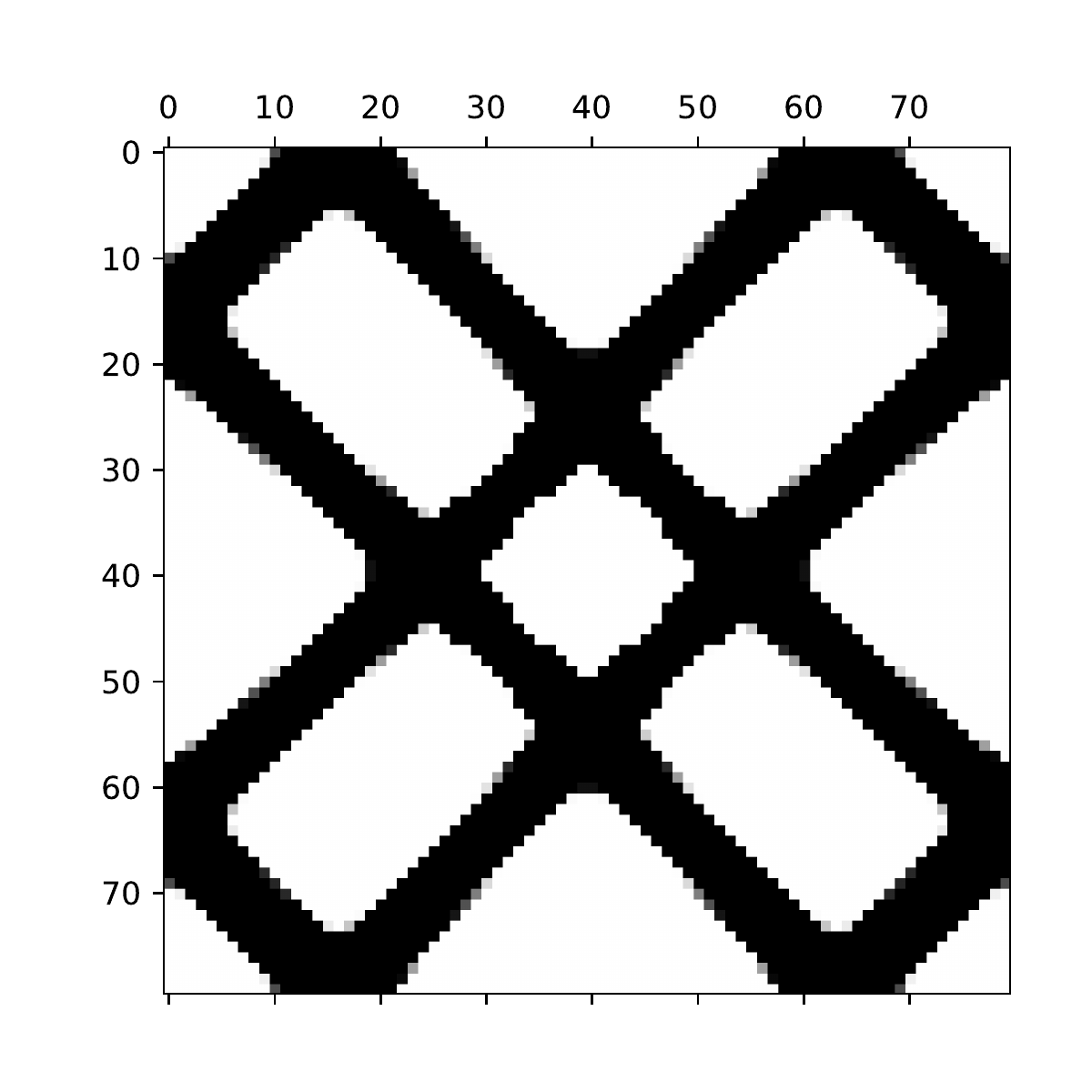}
        \label{fig:d8_}}
    \end{minipage}
    &
    \begin{minipage}[c]{0.285\textwidth}
       \centering 
        \subfloat[FEM, iteration 80, r=H/20]{\includegraphics[trim={1.6cm 0.5cm 0.5cm 1.6cm},clip,width=\textwidth]{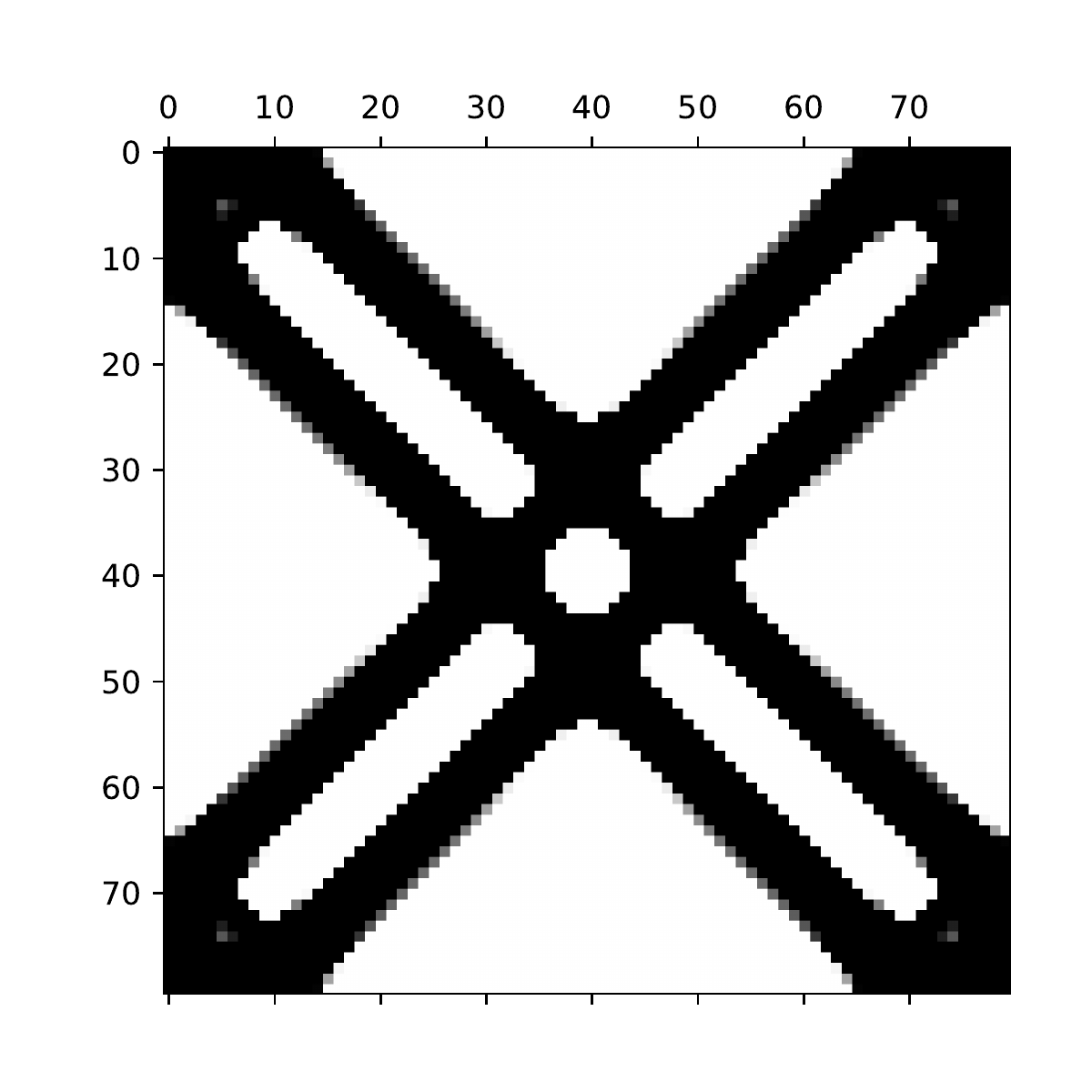}
        \label{fig:d9_}}
    \end{minipage}\\

    \end{tabular}
    \caption{Density plot of designs generated by DEM and FEM for 2D shear modulus maximization. The first row shows three different initial configurations, and the second and third rows show the corresponding final designs.}
    \label{2d_shear}
\end{figure}

In this example, a 2D unit cell of size 10-by-10 m$^2$ was subjected to a shear load at the top edge. The unit cell was subjected to periodic BCs in the X- and Y-direction. Further, 6561 nodes were placed into the domain forming a 81-by-81 grid. The objective is to maximize the homogenized shear modulus while maintaining the volume fraction at 40\%. It is well-known that the final optimized design depends heavily on the initial material distribution \cite{kollmann2020deep,zhang2019topology,xia2015design}. Therefore, three different initial designs were considered with different initial hole diameters, as shown in the first row of \fref{2d_shear}. $\epsilon_{tol}$ was set to $5\times10^{-6}$. For comparison, the DEM-based TO results are compared to their FEM-based counterparts produced by the MATLAB \cite{MATLAB:2021} code developed by Xia et al. \cite{xia2015design}.

\begin{figure}[h!] 
    \centering
     \subfloat[]{
         \includegraphics[width=0.33\textwidth]{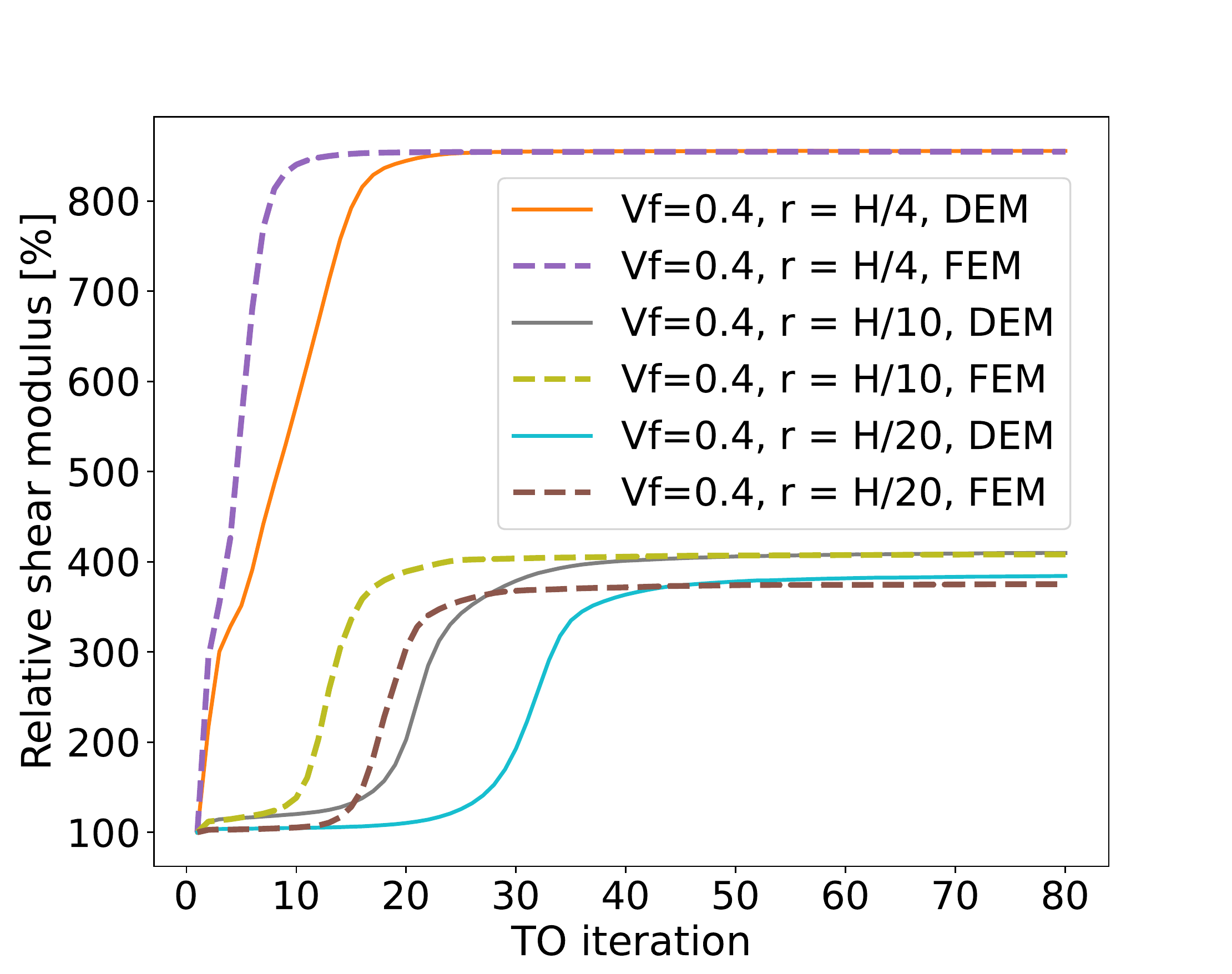}
         \label{fig:compliance_meta}
     }
     \subfloat[]{
         \includegraphics[width=0.33\textwidth]{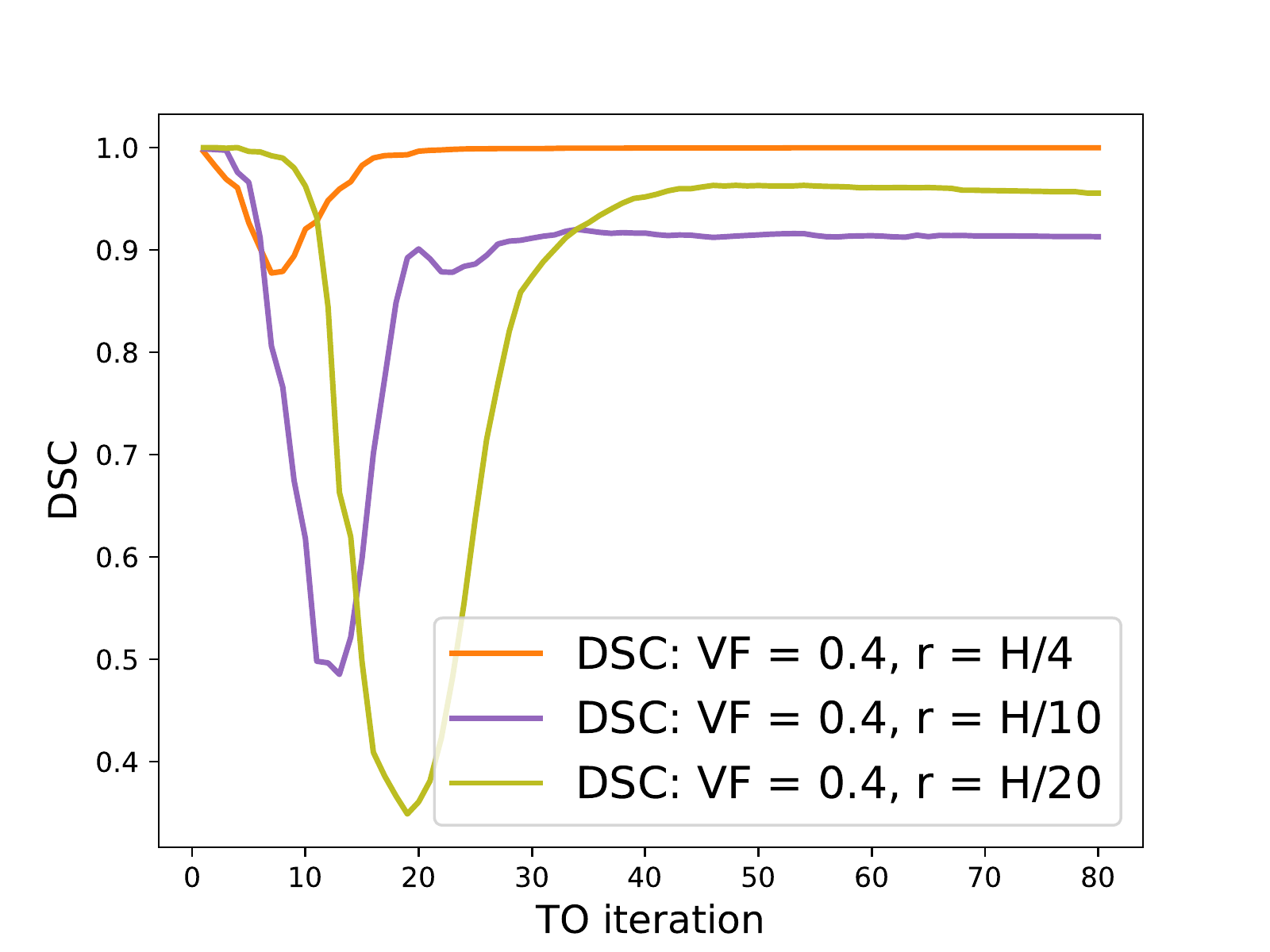}
         \label{fig:DSC_meta}
     }
     \subfloat[]{
         \includegraphics[width=0.33\textwidth]{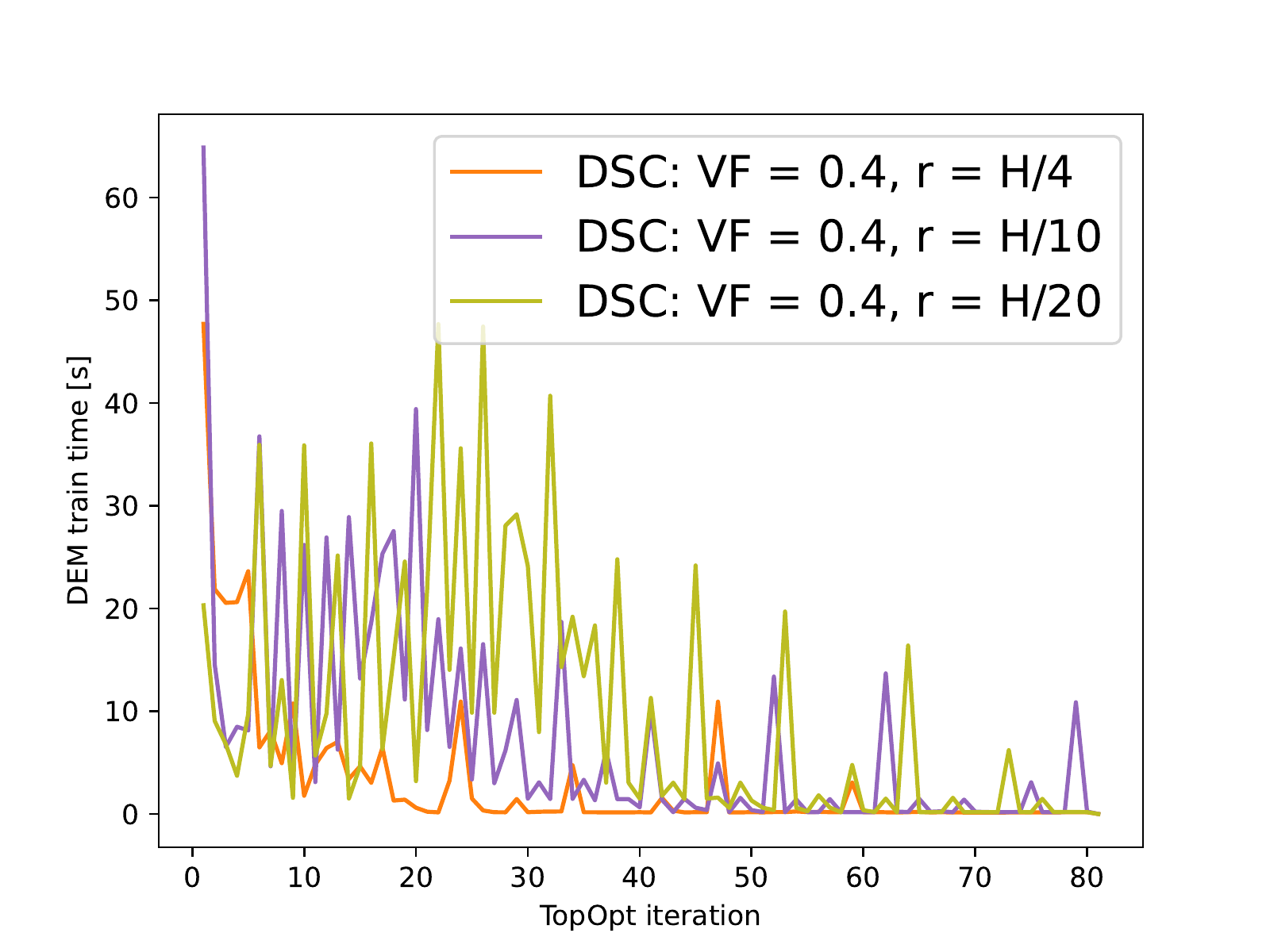}
         \label{fig:train_meta}
     }
    \caption{Comparing DEM-based TO with FEM-based TO in 2D shear modulus maximization: \psubref{fig:compliance_meta} Relative shear modulus evolution for the two cases. \psubref{fig:DSC_meta} Dice similarity coefficient computed on the binarized designs.
    \psubref{fig:train_meta} Training time for each iteration.}
\end{figure}

The initial and final designs from DEM- and FEM-based TO are compared in \fref{2d_shear}. The relative change in shear modulus for both cases over all the iterations are presented in \fref{fig:compliance_meta}. Further, DSC coefficient is calculated for the binarized final designs obtained from both cases using \eref{eq:DSC_eq} and is presented in \fref{fig:DSC_meta}. The DEM training time for the three cases were 579.7, 934.4 and 1125.7s, respectively. Training time is summarized in \fref{fig:train_meta}.

From \fref{2d_shear}, we clearly see that the DEM- and FEM-based TO frameworks produced very similar final designs for all three initial designs. However, comparing \fref{fig:d6_} and \fref{fig:d9_}, we noticed an important deficiency of the DEM-based TO method. The final FEM-based TO design (\fref{fig:d9_}) has four-fold symmetry, while the DEM-based design (\fref{fig:d6_}) does not. In fact, the DEM-based final design is not periodic at the lower left corner. This difference is due to that fact that periodic BCs are enforced in FEM-based simulations by constraining the corresponding degrees of freedom of the global stiffness matrix, hence periodicity of the displacement field (hence the material distribution) is also guaranteed. Also, if the initial design is symmetric, symmetry persists in subsequent designs. This is not the case for DEM-based TO, as the periodic BCs were enforced by penalty. The periodic penalty loss gradually creep up as the design progressed, and the designs gradually lost periodicity due to minor violations of periodicity in the displacement field and material distribution. Nonetheless, it is observed that the DSC is greater than 90\% for the final designs in all three cases. The low DSC during the initial stage of evolution is a result of the different optimizers used in the DEM- and FEM-based TO code. Similar to the two examples presented above, we again see gradual reduction of training time as TO iterations progressed.

\section{Conclusions and future work}
\label{sec:conc}
In this work, we present the use of DEM-based simulation technique in TO. The proposed framework combines solution to PDEs using PINNs and classical TO, where the momentum balance equation is solved by a deep neural network using DEM by minimizing the system potential energy. Instead of training a separate neural network to predict the updated density array as in the work of Zehnder et al. \cite{zehnder2021ntopo}, we leverage the fact that compliance minimization problem is self-adjoint, and express element sensitivity directly using the displacement field produced by the DEM model. Therefore, only one neural network needs to be trained in each optimization iteration, making the framework much simpler. The training of the DEM model also leverages transfer learning to reduce training time, where the weights and biases from the previous iteration are used as initial condition for the next iteration, which leads to significant reduction of training time in later iterations. The proposed framework combines two different types of optimizers, using a L-BFGS optimizer for the DEM model training, and using a classical MMA optimizer for updating the density array.

Three numerical examples are presented to demonstrate the capabilities of the proposed framework. In the first example, we performed compliance minimization in 2D on two different loading and boundary conditions, where we saw that the DEM-based topology optimization framework generated almost identical designs to its FEM counterpart. In the second example, we extended compliance minimization to 3D, where we optimized a 3D bridge under transverse loading. It was found that the hyperparameters used in 2D simulations were effective in 3D as well, generating a design that is quite similar to the one generated by Abaqus. In the last example, we attempted the design of 2D meta materials by maximizing the homogenized shear modulus, which necessitated the implementation of periodic boundary conditions in DEM. For three different initial designs, it was seen that our DEM framework was able to generate designs very similar to FEM, albeit suffered from minor loss of symmetry due to the inaccurate enforcement of periodic boundary conditions. In all examples presented, the training time for DEM model was longer than the solution time for FEM, but training time shows a gradual decrease with design iterations and is relatively insensitive to the number of nodes in the simulation domain. This makes the DEM-based TO method potentially beneficial when a fine mesh and many TO iterations are required.

We conclude that the proposed DEM-based topology optimization framework can produce optimized designs that are very similar to FEM. Although having a high computational cost, we emphasize that the goal of this work is not to outperform FEM in particular applications, but rather to highlight the diverse application scenarios for the DEM method, and shed light on how neural-network-based solutions for PDEs can be employed in engineering applications.

A primary limitation of this work is the need to form element discretization of the domain, and using shape function gradients to compute gradients of field variables produced by DEM, which is in contrast with many previous studies \cite{abueidda2021meshless,zehnder2021ntopo,abueidda2022deep} that used a meshless method to solve the underlying governing equation. Extension to a meshless formulation using automatic differentiation of the DEM model will be our future work. We will also explore topology optimization for hyperelastic materials undergoing finite deformation, where Newton-Raphson iterations need to be employed in FEM, rendering it rather expensive. In addition, the effects of enhancing the loss function with its gradient \cite{yu2022gradient,shukla2022scalable} on the DEM model accuracy will be studied.

\section*{Replication of results}
The data and source code that support the findings of this study can be found at: \url{https://github.com/Jasiuk-Research-Group}. \textcolor{red}{Note to editor and reviewers: the link above will be made public upon the publication of this manuscript. During the review period, the data and source code can be made available upon request to the corresponding author.}

\section*{Competing Interests}
The authors declare that they have no conflict of interest.

\section*{Acknowledgements}
The authors would like to thank the National Center for
Supercomputing Applications (NCSA) for providing the computational resources on the Delta cluster.

\section*{CRediT author contributions}
\textbf{Junyan He}: Conceptualization, Methodology, Software, Formal analysis, Investigation, Data Curation, Writing - Original Draft. \textbf{Shashank Kushwaha}: Software, Formal Analysis, Investigation, Writing - Original Draft. \textbf{Charul Chadha}: Software, Formal Analysis, Investigation, Writing - Original Draft. 
\textbf{Seid Koric}: Supervision, Resources, Writing - Review \& Editing.
\textbf{Diab Abueidda}: Supervision, Resources, Writing - Review \& Editing. \textbf{Iwona Jasiuk}: Supervision, Resources, Writing - Review \& Editing, Funding Acquisition.

\bibliographystyle{unsrtnat}
\setlength{\bibsep}{0.0pt}
{\scriptsize \bibliography{References.bib} }
\end{document}